\documentclass[journal,12pt,draftclsnofoot,onecolumn]{IEEEtran}

\usepackage{cite}
\usepackage{graphicx}
\usepackage{multicol}

\usepackage{amsfonts}
\usepackage{amssymb}
\usepackage{amsmath}
\usepackage{floatflt}

\begin{document}
\title{Yes-no Bloom filter: A way of representing sets with fewer false positives}
\author{Laura Carrea, Alexei Vernitski and Martin Reed%~\IEEEmembership{}% <-this % stops a space
%\thanks{%Manuscript received January 20, 2002; revised November 18, 2002.
%This work was supported by ....}% <-this
        %This work was supported by PURSUIT Project....}% <-this % stops a space
\thanks{Laura Carrea is affiliated with the School of Mathematical and Physical Sciences, University of Reading, UK and with the School of Computer Science and Electronic Engineering, University of Essex, UK; Alexei Vernitski is affiliated with the Department of Mathematical Sciences, University of Essex; Martin Reed is affiliated with School of Computer Science and Electronic Engineering, University of Essex, UK}}
\maketitle

\begin{abstract} The Bloom filter (BF) is a space efficient randomized data structure particularly suitable to represent a set supporting approximate membership queries. BFs have been extensively used in many applications especially in networking due to their simplicity and flexibility. The performances of BFs mainly depends on query overhead, space requirements and false positives. The aim of this paper is to focus on false positives. Inspired by the recent application of the BF in a novel multicast forwarding fabric for information centric networks, this paper proposes the yes-no BF, a new way of representing a set, based on the BF, but with significantly lower false positives and no false negatives. Although it requires slightly more processing at the stage of its formation, it offers the same processing requirements for membership queries as the BF. After introducing the yes-no BF, we show through simulations, that it has better false positive performance than the BF.  \end{abstract}


\section{Introduction}\label{intro}
The BF is a time and space efficient data structure that is used to represent concisely a set and allows highly efficient set membership queries \cite{bloom}. However, the BF has some probability of giving false positives in a membership test; that is, an element may appear to belong to the set when in fact it does not.
Moreover, elements can be added to the set, but not removed.
Because of their space-time efficiency, the BF has become very popular for network applications \cite{broder},  \cite{tarkoma}. 
In fact, BFs offer a representation which significantly reduces space and time to query at the cost of false positives. 
For many applications false positives can be tolerated or can be made low enough so that the savings in space and time they offer justify their use. However, finding methods to reduce false positives keeping the characteristics the BF offers is a challenge. 

Many variants of the BF have been proposed. Some of them, like the retouched BF proposed in \cite{donnet}, which is based on a randomized bit clearing, generates false negatives. Some other schemes allow the deletion of elements \cite{rothenberg2010:deletable} or counting the elements stored \cite{fan}, \cite{ficara}.  Almeida \cite{almeida:scalable} introduced the scalable BF, a BF variant that reduces false positives adapting dynamically to the number of elements stored, assuring a minimum false positive probability.  

Recently, the complement Bloom Filter \cite{lim15} has been proposed to reduce false positives. 
They introduce the complement Bloom filter which explicitly stores all the elements of the universe which are not in the set.  Unlike most variants of the Bloom filter, the complement Bloom filter produces false negatives, which are not tolerable in many applications. As is shown in the numerical experiment in \cite{lim15}, the size of the complement filter must be very large, making the construction difficult to implement and not space-efficient. Moreover, the complement filter is static: it does not have any parameters whose values can be chosen to improve the performance, unlike the construction proposed in this paper.
%They introduce the complement Bloom filter which stores all the elements of a universe which are not stored in the BF. However, firstly the data structure produces false negatives which are not tolerable in many applications. Secondly, the size of the complement filter have to be quite big so that the space efficiency is not necessarily preserved for this data structure. Finally, the structure is static and from the paper it is not clear how it would perform.  
Another structure which has been proposed to reduce false positives is the optihash \cite{carrea}. 
%if the set of element to query is known in advance. 
This variant has been inspired by the application of the BF as a forwarding identifier \cite{jokela} in the information centric network architecture proposed within the framework of the PSIRP/PURSUIT project \cite{fotiou}, \cite{trossen:pekka}. In this case, the BF becomes a fixed size forwarding identifier that is placed in the packet header and encodes a complete network path, or multicast tree, in a space/time efficient manner. This strategy was introduced to implement a forwarding layer replacing IP in applications such a clean-slate information centric architecture but it has been suggested also to be a good alternative to MPLS labeling \cite{zahemszkyMPSS}.
%False positives are an intrinsic problem of Bloom filters.

For the BF used as a network path encoding, a false positive means that traffic may be forwarded along links that were not intended to be part of the path. Thus, false positives may cause bandwidth wastage or may generate a loop, which can cause major problems in the network.  Hence, there is an interest in minimizing the false positive occurrences for this application and generally because of the BF wide applicability. 
For the application described in this paper, the total length is also an important parameter as the BF will occupy part of the packet header.  

In this paper, the \emph{yes-no} BF is presented. The aim of the yes-no BF is to reduce the number of false positives while keeping its properties of being a time and space efficient data structure. 
First, the classic BF is revised with a discussion on the false positive probability. Then, the yes-no BF structure in introduced and presented in details and the false positive probability is analytically evaluated. Also, an experimental analysis is carried out in order to evaluate the yes-no BF performance for false positives with respect to the yes-no BF parameters and the characteristics of the set which the structure represents. The average number of false positive occurrences of the yes-no BF is first analysed as a function of the set and data structure parameters and then it is evaluated for the new forwarding mechanism which has been proposed within the framework of the PSIRP/PURSUIT project. However, the data structure is general enough to be used in other applications. 
The complexity of the processing required at the stage of the formation of the yes-no BF is evaluated in terms of big O notation in comparison with the processing required for the classic BF.  
%Since the yes-no BF is used as the forwarding layer fabric, only the topology of a network will be of interest in the evaluation. Simulations will be carried out in realistic topologies.

\section{The Bloom filter}\label{BF}
The BF is a compressed way of representing a set of elements, with some false positives but no false negatives. A BF is a fixed length Boolean array where a certain number of bits are set to 1. 
Given a set $S$ of elements to store in the BF, each element of the set $S$ is represented with its own BF, a Boolean array of a fixed length $m$ where up to $k$ bits are set to 1. The bits are set using $k$ hash functions $\{h_1,h_2,...,h_k\}$ applied to each element $e_i$ of the set
\begin{equation}
    h_1(e_i), h_2(e_i),... h_k(e_i) \quad \text{with  } e_i \in E, \quad (i=1,...n),
\end{equation}  
where the output of each hash is in the range $\{1,2,...,m\}$. 

Given the BF $\textbf{b}_{e_i}$ of each element of the set $S$, the whole set is represented by a Boolean array, itself a BF, which can be interpreted as a logical disjunction of the BFs of its elements: 
\begin{equation}
   \textbf{b}_{S}= \textbf{b}_{e_1} \vee \textbf{b}_{e_1} ... \vee \textbf{b}_{e_n}.
\end{equation}

Given a set $T$ of elements to be queried, the membership test of an element in $T$ can be performed as a bitwise comparison between the BF of the set and the BF of the element. The comparison is traditionally implemented as a logical conjunction on $\textbf{b}_{S}$, the BF of the set and $\textbf{b}_e$, the BF of the element to be queried:
\begin{equation}
   \textbf{b}_{S} \wedge \textbf{b}_{e} \left\lbrace \begin{array}{l}
     = \textbf{b}_e \Rightarrow e\in S \\
     \neq \textbf{b}_e \Rightarrow e\not \in S
   \end{array} \right. ,  
\end{equation}
namely if the logical conjunction is equal to the BF of the element to query then the element necessarily belongs to the set otherwise it does not.
Another way to interpret the set membership query operation for the BF is to check bitwise if each bit of the BF of the element to query is less than or equal to the corresponding bit in the BF representing the set. This is easy to verify using the truth table of the logical conjunction reported in Table \ref{table_example}, where $\textbf{b}_e\wedge \textbf{b}_S = \textbf{b}_e$, then $\textbf{b}_e \leq \textbf{b}_S$ bitwise.
\begin{table}[tb]
\renewcommand{\arraystretch}{1.6}
\caption{True table of a logical conjunction.}
\label{table_example}
\centering
\begin{tabular}{c|c|c}
$\textbf{b}_e$ & $\textbf{b}_S$ & $\textbf{b}_e\wedge \textbf{b}_S$ \\
\hline
 0 & 0 & 0 \\
 0 & 1 & 0  \\
 1 & 0 & 0 \\
 1 & 1 & 1
\end{tabular}
\end{table}

%In hardware this can performed using a straightforward bitwise AND and compare.

Because of the probabilistic nature of the BF, the BF can be only queried and its elements can not be deleted. Moreover, the set membership test may give rise to a false positive: \emph{i.e.} an element appears to belong to the set even though it was not originally included in the BF. 
Let us consider a BF of $m$ bits. The probability that a specific bit is still 0 after all 
the elements in $S$ have been hashed with $k$ hash functions is \cite{tarkoma}:
\begin{equation}
    p = \left(1-\frac{1}{m} \right)^{kn} 
\end{equation}
where $n=|S|$ and where it has been assumed that the hash functions are perfectly random.

\begin{figure}
	\centering
	\includegraphics[width=2.0in]{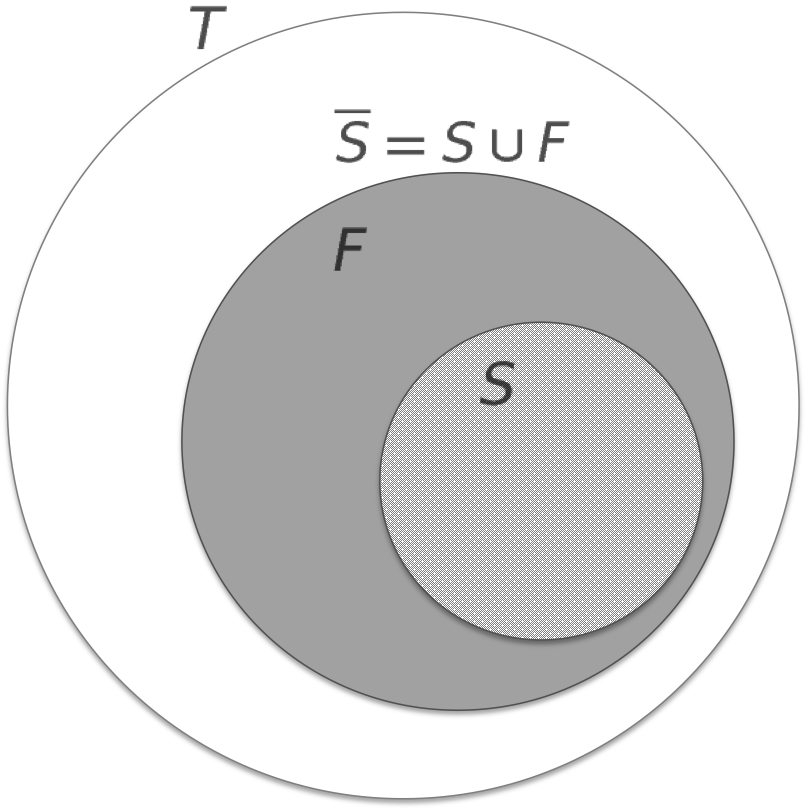}
	\caption{Set diagram for the elements included in the BF, the false positives and the elements to be queried.}
	\label{BF_set}
\end{figure}

Given the BF $\textbf{b}_S$ of the set $S$ and the set $T$ of elements to be queried, let $\overline{S}$ the set of elements $e$ such that $\textbf{b}_e \le \textbf{b}_E$, namely the set of elements that appears to belong to the set $S$ after a membership query. We can call $\overline{S}$ the set of the positives of the BF and $S \subset \overline{S}$ since the elements in $S$ definitely satisfy the condition. Then, the set $F=\overline{S}\setminus S$ is the set of the false positives. The sets are schematically represented in Fig. \ref{BF_set}, where $S \cap F = \varnothing$.

For an element $e \not \in S$, the probability that it belongs to the BF (namely that all the bits $h_1(e), h_2(e), ... h_k(e)$ are set to 1) when tested for membership can be computed as \cite{tarkoma}
\begin{equation}\label{f_S}
    f_S = (1-p)^k = \left(1- \left(1-\frac{1}{m} \right)^{kn} \right)^k,
\end{equation}
where the assumption that if $B_i$ and $B_j$ are two distinct bits in the Bloom filter, the events $B_i = 1$ and $B_j = 1$ are statistically independent has been made.
However, this is not necessary true as shown in \cite{bose}. 
Consequently, (\ref{f_S}) is only an approximation of the false positive probability. In particular, Bose in \cite{bose} showed that (\ref{f_S}) a strict lower bound for any $k \geq 2$ and proposed a calculation of the false positive probability which is not though in closed form. 
A further approximation of (\ref{f_S}) is
\begin{equation}\label{f_Se}
    f_S \approx \left(1-\text{e}^{-\frac{kn}{m}} \right)^k
\end{equation}
being $\text{e}^x = \lim_{a \rightarrow \infty} \left(1+\frac{x}{a}\right)^a$.

This probability is computed assuming that $e\not \in S$, namely it is the conditional probability \cite{kirsch}
\begin{equation}\label{fs_prob}
     f_s = \text{Pr}[\overline{S}|S^c]
\end{equation}
where $S^c = T \setminus S$ is the complement of $S$.

Instead, the probability of a false positive without assuming $e \in S^c$ can be written as \cite{lim}
\begin{equation}
    \text{Pr}[F] = \text{Pr}[\overline{S} \setminus S] = \text{Pr}[\overline{S}] - \text{Pr}[S],
\end{equation}  
since $\overline{S}=S \cup U$ and $S \cap U = \varnothing$.
The probability of $\overline{S}$ can be expressed as:
\begin{equation}
     \text{Pr} [\overline{S}] = \text{Pr}[S] \,\, \text{Pr} [\overline{S}|S]+\text{Pr}[S^c]\,\, \text{Pr} [\overline{S}|S^c].
\end{equation}
Since $\text{Pr}[\overline{S}|S]=1$, $\text{Pr} [\overline{S}|S^c]=f_S$ from (\ref{fs_prob}) and $\text{Pr} [S^c]=1-\text{Pr}[S]$, we obtain 
\begin{equation}\label{PrF(S)}
    \text{Pr} [\overline{S}] = \text{Pr}[S] + (1-\text{Pr}[S]) f_S
\end{equation}
so that the probability of false positive for $e \in T$ can be written as
\begin{equation}\label{PrF}
   \text{Pr}[F] = \text{Pr}[\overline{S} \setminus S] = (1-\text{Pr}[S]) f_S.
\end{equation}
If the condition $e \in S^c$ is considered, namely $e \not \in S$ then $\text{Pr}[S] =0$ and (\ref{PrF}) becomes (\ref{fs_prob}).

The advantages of the BF are that the calculations needed to build it, and test for membership, are easy to program and very fast to perform, since only simple logical operations are required. A limit of the BF is that its probabilistic nature gives rise to false positives. The aim of this work is to propose a mechanism to limit the number of false positives.

\section{The yes-no BF} \label{yesno}
The \emph{yes-no BF} we propose is a new way of representing a set and it is based on the BF. It is composed by two parts:
\begin{itemize}
   \item the \emph{yes-filter} encoding the set;
   \item the \emph{no-filter} encoding the elements which generate false-positives.
\end{itemize}

It aims to offer a smaller number of false positives when compared to the BF and it can be designed to have no false negatives. The main insight behind the yes-no BF is that, in many applications, not only the elements belonging to the set but, more generally, all the elements which will ever be queried are known when the data structure is formed. 
%We need to point out that there are applications of the BF where the set of elements to be tested are not necessarily known in advance. However, for the yes-no BF knowing all elements in advance is a requirement.

Let $U$ be the universe of all elements, $S\subseteq U$ be the set of elements to be encoded in the yes-no BF, and $T\subseteq U \setminus S$ be the set of all the elements which do not belong to the set $S$ and whose membership in $S$ is likely to be tested. Typically, $T$ is a smaller set than $U\setminus S$.

The yes-no BF we propose, consists of a structure that not only has the function to encode the elements of $S$, it has also some additional features to avoid actively the false positives during the set membership test.  Moreover, the yes-no BF aims to preserve the space-time efficiency feature offered by the BF, although the structure requires slightly more processing but only at the stage of its formation.  

\begin{figure}
  \centering
  \includegraphics[width=3.0in]{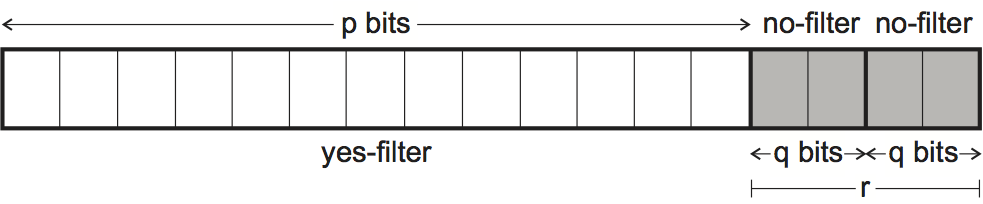}
  \caption{Schematic representation of the yes-no BF structure.}
  \label{YesNo}
\end{figure}

As with the standard BF, we assume that the representation of a set consists, in total, of $m$ bits. These $m$ bits are split into the following parts: \begin{itemize}
  \item $p$ bits, to which we shall refer as the \emph{yes-filter};
  \item $qr$ bits, where we shall refer to each of the $q$ bits as to a \emph{no-filter}.
\end{itemize}
so that $m=p+qr$ and $q<<p$.
We shall refer to the above construction as the \emph{yes-no BF} and a schematic representation is shown in Fig. \ref{YesNo}.

\subsection{Representation of an element in $U$}\label{yesnoele}
Before describing the set construction we will introduce the construction of a single element which will have a yes-filter and a single no-filter repeated $r$ times. Two different sets of hash functions are needed to represent each element $e \in U$ with a yes-no BF structure: 
\begin{itemize}
	\item $H=\{h_1,h_2,...h_k\}$ for the yes-filter
	\item $W=\{w_1,w_2,..,w_{k'}\}$ for the no-filter
\end{itemize}
where generally $k'<k$. Each hash function in $H$ outputs values in $\{1,2,..p\}$ while each hash function in $W$ outputs values in $\{1,2,..q\}$, with $q<<p$ and $qr<p$. 

To construct the yes-filter the element $e\! \in \! S$ is hashed $k$ times and the bits $h_i(e)$ in the yes-filter are set to 1. To construct the no-filter the element $e$ is hashed $k'$ times and the bits $w_j(e)$ in the no-filter are set to 1.  Note that each element $e$ has only one no-filter. The yes-no BF representation of the element $e$ is constructed with its yes-filter and  with $r$ repetitions of the no-filter. For convenience, we shall refer to these parts as the yes-filter and the no-filter of $e$.

\begin{figure}[h]
  \centering
  \includegraphics[width=3.0in]{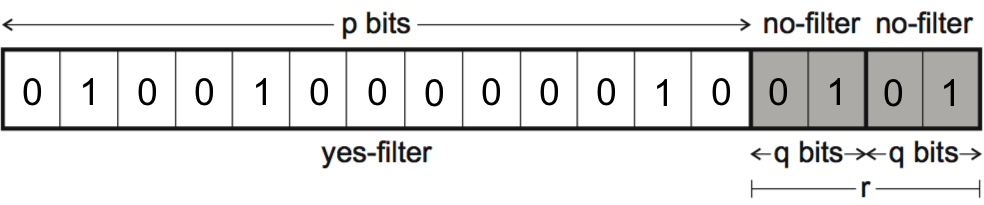}
  \caption{An example of the yes-no BF structure for an element $e$.}
  \label{YesNoExample}
\end{figure}

As an example, we consider $p=13$, $q=2$, $r=2$ as shown in Fig. \ref{YesNoExample}. The number of hash functions considered are $k = 3$ and $k'=1$, and $h_1(e)=5$, $h_2(e)=2$, $h_3(e) = 12$ and $w_1(e)=2$. 

\subsection{Representation of a set in $U$}\label{yesnoset}
Given the set $S$ of elements to encode in the yes-no BF and the set $T\subseteq U\setminus S$ of elements whose membership in $S$ may be queried, the yes-no BF representing the set is constructed using the yes-filters of each element of the set $S$ and the no-filters of the elements in $T$ which produce false positives when the yes-filter is queried. In particular, the yes-filter and the no-filters of the set are constructed as described below.

\subsubsection{The yes-filter}
The yes-filter of the set $S$ is formed as a disjunction of the yes-filters of the all the elements $e \in S$, exactly as for a normal BF.

\subsubsection{The no-filter}\label{nofilter}
The no-filters of the set store the no-filters of the elements which generate false positives when the yes-filter is queried. The yes-no filter reduces false positives compared to the BF by keeping track, in an efficient way, not only of the elements belonging to the set $S$ but also of the elements which generate false positives.

Let us analyse in detail the procedure to construct the no-filters of the set $S$.
Firstly, the set $F \subseteq T$ of elements $f$ which generate false positives is constructed using the membership query procedure described for the BF (see Section \ref{BF}). Namely, the yes-filter of each of the elements in $T$ is checked against the yes-filter of the set. That is, if bit-wise each bit of the yes-filter of an element in $T$ is less than or equal to the yes-filter of the set $S$ then necessarily this element generates false positives and it will be an element of the subset $F \subseteq T$. %it is stored in the set $F$. 
Generally, $|F|>r$, namely the number of false positives is greater than the number of no-filters. Consequently, each of the no-filters of the set $S$ will have to store more than one no-filter. 

\begin{figure}[h]
	\centering
	\includegraphics[width=2.5in]{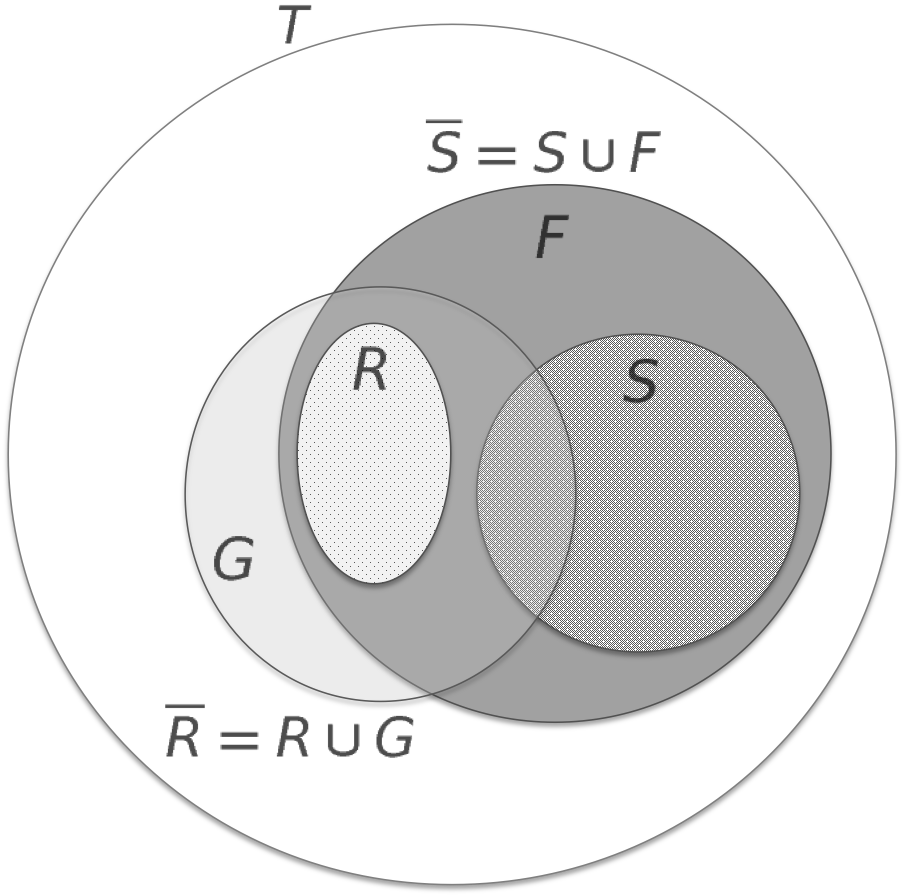}
	\caption{Set diagram of the elements to include in the yes- and no-filters, the false positives of the yes- and the no-filters and the elements to be queried for the case with false negatives.}
	\label{YN_set}
\end{figure}

Each no-filter of the set is a BF which, obviously, may also give rise to false positives. A false positive of a no-filter would mean that the element was not in reality a false positive for the yes-filter even though the no-filter represents it as a false positive of the yes-filter.
An element which is not a false positive of the yes-filter may be an element which was added to the yes-filter. Consequently, a false positive of the no-filter may represent a false negative of the yes-filter. Fig. \ref{YN_set} shows the set $R$ of elements that have been included in the no-filters, namely $R \subset F$. Also, it shows the set $G$ of the false positives of the no-filters which may have some elements in $S$, namely the set $G \cap S$ is the set of the false negatives.

Hence, in the construction it may be required to avoid false negatives in the yes-filter since such false negatives may be more harmful\footnote{for some applications like forwarding in information centric networks.} than false positives.  The following steps are proposed for the construction of the no-filters of the set $S$ in order to avoid false negatives, but the design can be altered if false negative can be tolerated. 

\begin{figure}[h]
	\centering
	\includegraphics[width=2.5in]{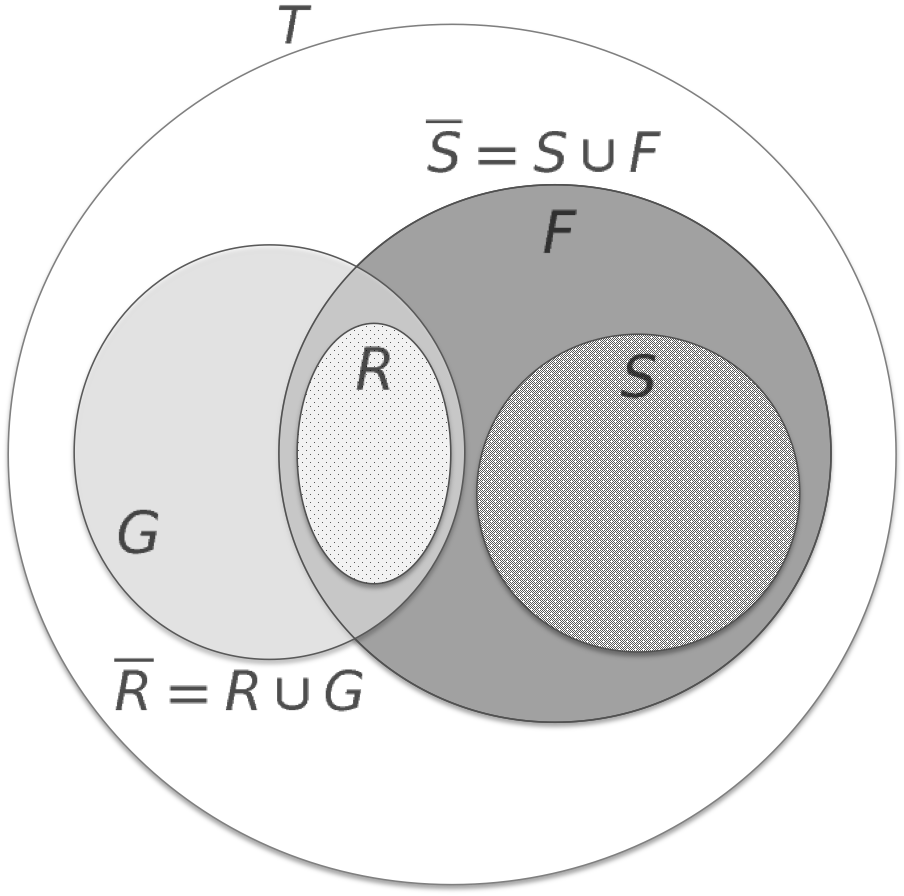}
	\caption{Set diagram of the elements to include in the yes- and no-filters, the false positives of the yes- and the no-filters and the elements to be queried for the case with no false negatives.}
	\label{YN_set_noFN}
\end{figure}

For each element in $F$, the algorithm will try in turn each no-filter stored in the yes-no filter of the set to test if it does create a false negative. However, if none are suitable, then, this particular false positive cannot be mitigated. Let us see in detail this construction. 
\begin{itemize}
	\item Given the no-filter $\textbf{\texttt{v}}_{f_1}$ of the first false positive $f_1 \in F_{yes}$ and the first no-filter $\textbf{\texttt{v}}_1$ of the set S, a disjunction $\textbf{\texttt{v}}_1'$ is formed is formed between them to attempt to store the no-filter of the first false positive in the yes-no BF: 
	\begin{equation}
	    \textbf{\texttt{v}}_1'=\textbf{\texttt{v}}_{f_1} \vee \textbf{\texttt{v}}_1.
	\end{equation}
	\item Then, to avoid false negatives, the following condition is checked for the no-filter of each element $e \in S$:
		\begin{itemize}
			\item if the no-filter $\textbf{v}_e$ of each $e \!\in\! S$ is greater than $\textbf{\texttt{v}}_1'$, then none of the no-filters of the elements in $S$ is a false positive for the no-filter storing the false positives, namely none of the elements in $S$ can be a false negative. In this case, the first no-filter of $S$ will be set as $\textbf{\texttt{v}}_1'$.
			\item if the no-filter $\textbf{v}_e$ of any $e \in S$ is less than or equal to $\textbf{\texttt{v}}_1'$ then it may give raise to false negative. In this case, the second no-filter $\textbf{\texttt{v}}_2$ of the set $S$ is attempted to be used following the same steps as for the first no-filter. If also the last filter $\textbf{\texttt{v}}_r$ satisfy this condition then, $f_1$ cannot be included in any of the no-filters and necessarily will be a false positive.
		\end{itemize}
\end{itemize}
This process is repeated for all the element in $F$ and the set $R$ of elements to store in the no-filters is constructed.
For convenience, we shall refer to the parts of the yes-no BF of the set $S$ as the yes-filter and the no-filters of the set $S$ with respect to the set $T$. Fig. \ref{YN_set_noFN} shows the sets when false negatives are avoided. In this case, $G \cap S = \varnothing$. 

We need to point out that 
%apart from the choice of the number $r$ of no-filters and their sizes $q$, 
the no-filter for each false positive occurrence can be placed in anyone of the $r$ no-filters of the set $S$. One would aim to choose the no-filters of the set $S$ in such a way that between them, they minimise the number of false positives. However, this is a complex discrete optimisation problem; it is not unreasonable to conjecture that finding an exact solution to it (that is, the smallest possible number of false positives, given a fixed number of no-filters) is NP-hard. 
The algorithm described above is a simple and fast greedy algorithm, and we show in the experiments below that
its performance provides better false positive rates than the BF. However, in some implementations more time-consuming algorithms may be used, which use the arsenal of operational research methods to further reduce false positive rates of yes-no Bloom filters. Optimisation techniques to form the no-filters in a near-optimal way have been proposed in \cite{yang}.

%Optimisation techniques to form the no-filters in a near optimal way have been proposed in \cite{yang}. However, for our model to be workable, we cannot afford to use time-consuming algorithms; consequently, we suggest to use a simple and fast greedy algorithm, and we show in the experiments below that its performance provides better false positive rates than the BF.

\subsection{Membership queries}\label{yesnoquery}
Given an element $t \in T$, to perform a membership query the following steps are required.
\begin{itemize}
\item Firstly, the yes-filter of $t$ is compared with the yes-filter of the set. If the yes-filter of $t$ is greater than the yes-filter of the set, the element $t$ definitely does not belong to the set $S$. In case that the yes-filter of $t$ is less than or equal to the yes-filter of the set, the element $t$ may be either an element belonging to $S$, or a false positive. Thus, in this case, the following step is performed.
	\item The no-filters of the set $S$ are checked. If the no-filter of $t$ is less than or equal to any of the no-filters of the set then we can conclude that $t$ was a false positive and therefore, discarded. Otherwise, we assume that $t$ was not a false positive of the yes-filter and that it is an element of $S$.
\end{itemize}

To conclude, the yes-no BF is a generalisation of the standard BF: indeed, when $r=0$, the yes-no BF turns into the BF of length $m$. Keeping $m$ fixed, when $r>0$, the length of the yes-filter is reduced (compared with the case $r=0$), producing an increased number of false positives with high probability; however, at the same time, the no-filters reduce the number of false positives as shown in the next section. The yes-no BF offers no false negatives. 

\section{False positive probability of the yes-no BF}

Let us consider the case of yes-no BF with no false negatives in Fig. \ref{YN_set_noFN}. The set $S$ is the set of elements to store in the yes-filter and the set $\overline{S}$ is the set of all the elements that, when tested, appear to belong to the yes-filter. Consequently, $\overline{S}\setminus S = F$ is the set of the false positives of the yes-filter. 

The set $R$ is the set of elements that are stored in the no-filter after the formation processing described in Sec. \ref{yesnoset}. The set $R \subset F$ is the set of elements generating false positives for the yes-filter which are stored in the no-filters. 
The set $\overline{R}$ is the set of all the elements that, when tested, appear to belong to the no-filter. Consequently, $\overline{R}\setminus R = G$ is the set of the false positives of the no-filter. In this case, $G$ and $S$ are disjoint. 

%To compute the false positive probability, let us consider 
If $e \in \overline{S}^{\,c}$ then $e$ is discarded by the yes-filter and consequently it is not queried by the no-filters.
If $e \in \overline{S}$ than $e$ is queried by the no-filter. The set $\overline{S}$ can be decomposed in the following 4 sets (see Fig. \ref{YN_set_noFN}):
\begin{itemize}
	\item $S$
	\item $R$
	\item $(\overline{R}\setminus R) \cap \overline{S}$
	\item $\overline{S}\setminus S \setminus \overline{R}$
\end{itemize}

If $e \in S$, $e$ belongs to the set $S$ of elements that were stored in the yes-no BF.

If $e \in R$, then $e$ was a false positive that has been discarded by the no-filters.

If $e \in (\overline{R}\setminus R) \cap \overline{S}=G \cap F$ then $e$ was a false positive of the no-filter but it was also a false positive  of the yes-filter so it was discarded by the no-filter.

If $e \in \overline{S}\setminus S \setminus \overline{R}=F\setminus \overline{R}$ then necessarily $e$ is a false positive of the yes-no BF.

We call the set $\overline{S}\setminus S \setminus \overline{R}=F\setminus \overline{R}$ the set $E$.   

\textbf{Lemma.} \textit{The probability of the set $E$ can be formulated as}
\begin{eqnarray}
&& \text{Pr}[E] =(1-\text{Pr}[S])f_S+ \\
&& -(\text{Pr}[S]+(1-\text{Pr}[S])f_S)f_R-(1-f_R)\text{Pr}[R]. \nonumber
\end{eqnarray}

\begin{proof}

The set $\overline{S} \setminus S$ is the union of the following disjoint sets:
\begin{equation}
    \overline{S} \setminus S = [\overline{S}\setminus S\setminus \overline{R}] \cup [(\overline{R}\setminus R) \cap \overline{S}] \cup R
\end{equation}
so that the probability $\text{Pr}[E]=\text{Pr}[\overline{S}\setminus S\setminus \overline{R}]$ can be written as
\begin{equation}\label{PrE}
    \text{Pr}[E] = \text{Pr}[\overline{S} \setminus S] - \text{Pr}[(\overline{R}\setminus R) \cap \overline{S}] -\text{Pr}[R].
\end{equation}
Now, $\text{Pr}[\overline{S} \setminus S]$ can be calculated as in (\ref{PrF}):
\begin{equation}\label{PrE2}
     \text{Pr}[\overline{S} \setminus S] = (1-\text{Pr}[S])f_{S}. 
\end{equation}
The probability $\text{Pr}[(\overline{R}\setminus R) \cap \overline{S}]$ can be written as
\begin{equation}\label{Pr3}
     \text{Pr}[(\overline{R}\setminus R) \cap \overline{S}] = \text{Pr}[(\overline{R}\setminus R)] - \text{Pr}[\overline{R}\setminus \overline{S}]
\end{equation}
since the set $\overline{R}\setminus R$ is the union of the two disjoint sets $(\overline{R}\setminus R) \cap \overline{S}$ and $\overline{R}\setminus \overline{S}$ as can be seen in Fig. \ref{YN_set_noFN}.
Now, $\text{Pr}[(\overline{R}\setminus R)]$ can be computed as in (\ref{PrF}):
\begin{equation}\label{PrFn}
    \text{Pr}[(\overline{R}\setminus R)] = (1-\text{Pr}[R]) f_{R}.
\end{equation}
while the probability $\text{Pr}[\overline{R}\setminus \overline{S}]$ can be computed as
\begin{eqnarray}\label{Pr1}
    && \text{Pr}[\overline{R}\setminus \overline{S}] = \text{Pr}[\overline{R} \cap \overline{S}^c]= \nonumber \\
    && = \text{Pr}[\overline{S}^{\,c}] \,\, \text{Pr}[\overline{R}|\overline{S}^{\,c}].
\end{eqnarray}
Regarding $\text{Pr}[\overline{R}|\overline{S}^{\,c}]$, since $\overline{S}^{\,c} \subset R^c$, namely if $e \in \overline{S}^{\,c}$, then necessarily $e \in R^c$, we can write
\begin{equation}\label{PrFSnSy}
    \text{Pr}[\overline{R}|\overline{S}^{\,c}] = \text{Pr}[\overline{R}|R^c] = f_{R}.
\end{equation}
Regarding $\text{Pr}[\overline{S}^{\,c}]$ it can be expressed using (\ref{PrF(S)}) as:
\begin{equation}\label{PrFSyc}
    \text{Pr}[\overline{S}^{\,c}] = 1-\text{Pr}[\overline{S}] = 1 - (\text{Pr}[S]+(1-\text{Pr}[S])f_{S}.
\end{equation}
Using (\ref{PrFSyc}) and (\ref{PrFSnSy}) we can rewrite (\ref{Pr1}) as
\begin{equation}\label{Pr2}
    \text{Pr}[\overline{R}\setminus \overline{S}] =  f_{R} (1 - (\text{Pr}[S]+(1-\text{Pr}[S])f_{S})).
\end{equation}
Using (\ref{PrFn}) and (\ref{Pr2}) we can rewrite (\ref{Pr3}) as
\begin{eqnarray}\label{PrE1}
    &&\text{Pr}[(\overline{R}\setminus R) \cap \overline{S}] =  (1-\text{Pr}[R])f_{R} + \nonumber \\
    && - (1 - (\text{Pr}[S]+(1-\text{Pr}[S])f_{S})) f_{R} = \nonumber \\
    && = (\text{Pr}[S]-\text{Pr}[R] +f_{S}-\text{Pr}[S]f_{S} ) f_{R}. 
\end{eqnarray}
From (\ref{PrE1}) and (\ref{PrE2}) we can compute the probability of the false positives of the yes-no BF $\text{Pr}[E]$ defined in (\ref{PrE}) as
\begin{eqnarray}\label{PEfinal}
    && \text{Pr}[E] =(1-\text{Pr}[S])f_S+ \\
    && -(\text{Pr}[S]+(1-\text{Pr}[S])f_S)f_R-(1-f_R)\text{Pr}[R] \nonumber
\end{eqnarray}
\end{proof}

The probability $\text{Pr}[E]$ in (\ref{PEfinal}) is the false positive probability for $e \in T$ for the yes-no BF. It is very easy to verify that the probability of false positive of the yes-no BF is less than the correspondent probability of false positive of a BF of the same size as the yes-filter:
\begin{equation}\label{PE<PF}
    \text{Pr}(E) < \text{Pr}(F)
\end{equation}
where $\text{Pr}(F)$ is defined in (\ref{PrF}). 
This probability has been computed without assuming that $e \in S^{\, c}$.
Instead, with the assumption that $e \in S^{\, c}$, then the probability of false positives can be written as
\begin{equation}
   \text{Pr}[E|S^c] = \frac{\text{Pr}[E \cap S^c]}{\text{Pr}[S^c]}.
\end{equation}
Since $E \subset S^c$, then $\text{Pr}[E \cap S^c] = \text{Pr}[E]$.
Consequently, given $e \in S^c$, $\text{Pr}[S^c]=1$, $\text{Pr}[S]=0$ and the false positive probability in (\ref{PEfinal}) becomes:
\begin{equation}
    \text{Pr}[E|S^c] = f_S(1- f_R) - (1-f_R)\text{Pr}[R].
\end{equation}
If we consider $e \in S^c \cap R^c$, assuming that the no-filter has taken care of the false positives, then $\text{Pr}[R]=0$ and the false positive probability becomes:
\begin{equation}\label{fyesno}
     \text{Pr}[E|S^c\cap R^c] = f_S(1- f_R).
\end{equation}
Thus, assuming that $e \in S^c\cap R^c$, the false positive probability of the yes-no BF depends on the false positive probability of the yes-filter $f_S$ and on the probability $1-f_R$ for the no-filters, where $f_R$ is the probability of false positives of the set of no-filters. 
It is easy to verify that the probability of false positives of the yes-no BF $\text{Pr}[E|S^c\cap R^c]$ is smaller than the probability of false positives of a Bloom filter of the same size of the yes-filter.   

If only one no-filter is employed, the false positive probability can be explicitly written in function of the yes-no BF parameters. The false positive probability of the yes-filter can be approximated using (\ref{f_Se}):
\begin{equation}
    f_S \approx \left(1-\text{e}^{-\frac{kn}{p}}\right)^k.
\end{equation}
The false positive probability $f_R$ of the no-filter can be approximated as
\begin{equation}
f_R \approx \left(1-\text{e}^{-\frac{k'n}{q}}\right)^{k'}.
\end{equation}
The false positive probability of the yes-no BF $f_E$ in  (\ref{fyesno}) with one no-filter only could be approximated as
\begin{equation}
    f_E = \left(1-\text{e}^{-\frac{kn}{p}}\right)^k \left(1- \left(1-\text{e}^{-\frac{k'n}{q}}\right)^{k'} \right).
\end{equation}

However, since the choice of how to fill the no-filters is related to a NP-hard problem (see end of Sec. \ref{nofilter}) it is not possible to express $f_R$ for a number of no-filters.
%%%%%%%%%%%%%%%%
For this reason, since choices need to be made to implement yes-no BFs, simulations has been carried out to study the dependency of the number of false positive from the design parameters. As with the standard BF, one would expect that the parameters defining the structure of the yes-no BF  have to be decided upon in advance. The parameters that need to be selected are: $m$, $r$, $p$, $q$ and the values of $k$ and $k'$ for the yes-filter and for the no-filters respectively.  

\section{Evaluation of the yes-no BF performance with respect to the yes-no BF parameters} \label{evalFP}

An experimental evaluation has been carried out in order to study the dependency of the false positive rate from the yes-no BF parameters and from the set $S$ characteristics. 
%Regarding a potential analytical expression of the false positive rate as a function of the yes-no BF parameters, the problem is very complex and most likely it will not be possible to find an analytical expression in closed form. The exact false positive probability for the Bloom filter itself which is a much simpler structure already cannot be expressed in a closed form as shown in \cite{bose}.

The number of false positives for the yes-no BF has been quantified through simulations and plotted against the average number of false positives of the correspondent BF calculated using the approximate formula of the false positive probability $f_p$ in (\ref{f_S}).
Given the false positive probability, the average number $F_p$ of false positive occurrences can be computed as
\begin{equation}
    F_p = \sum_{t\in T} \, f_p = \, \mid T\! \mid f_p
\end{equation} 
since the probability is the same for each element in $T$.

The evaluation of the false positive performance has been carried out as function of the cardinality of $S$ and as a function of the yes-no BF parameters. The following values have been used: 
\begin{itemize}
	\item $m=256$, the total number of bits of the yes-no BF. This is a typical size for in-packet forwarding applications \cite{jokela}
	\item $n=30$, the number of elements to store
	\item $t=100$, the number of elements to query
	\item $p=160$, the length of the yes-filter
	\item $q=32$, the length of a no-filter
	\item so $r=3$, the number of no-filters 
	\item $k= 4 $, the number of hash functions for the yes-filter
	\item $k'=5$, the number of hash function for the no-filter
\end{itemize}

Depending on the functionality studied each of these parameter has been varied across a specific range as described in the following.

\subsection{Evaluation of false positive performance as a function of the number of hash functions for the yes-no BF}

\begin{figure}[h]
	\centering
	\includegraphics[width=3.8in]{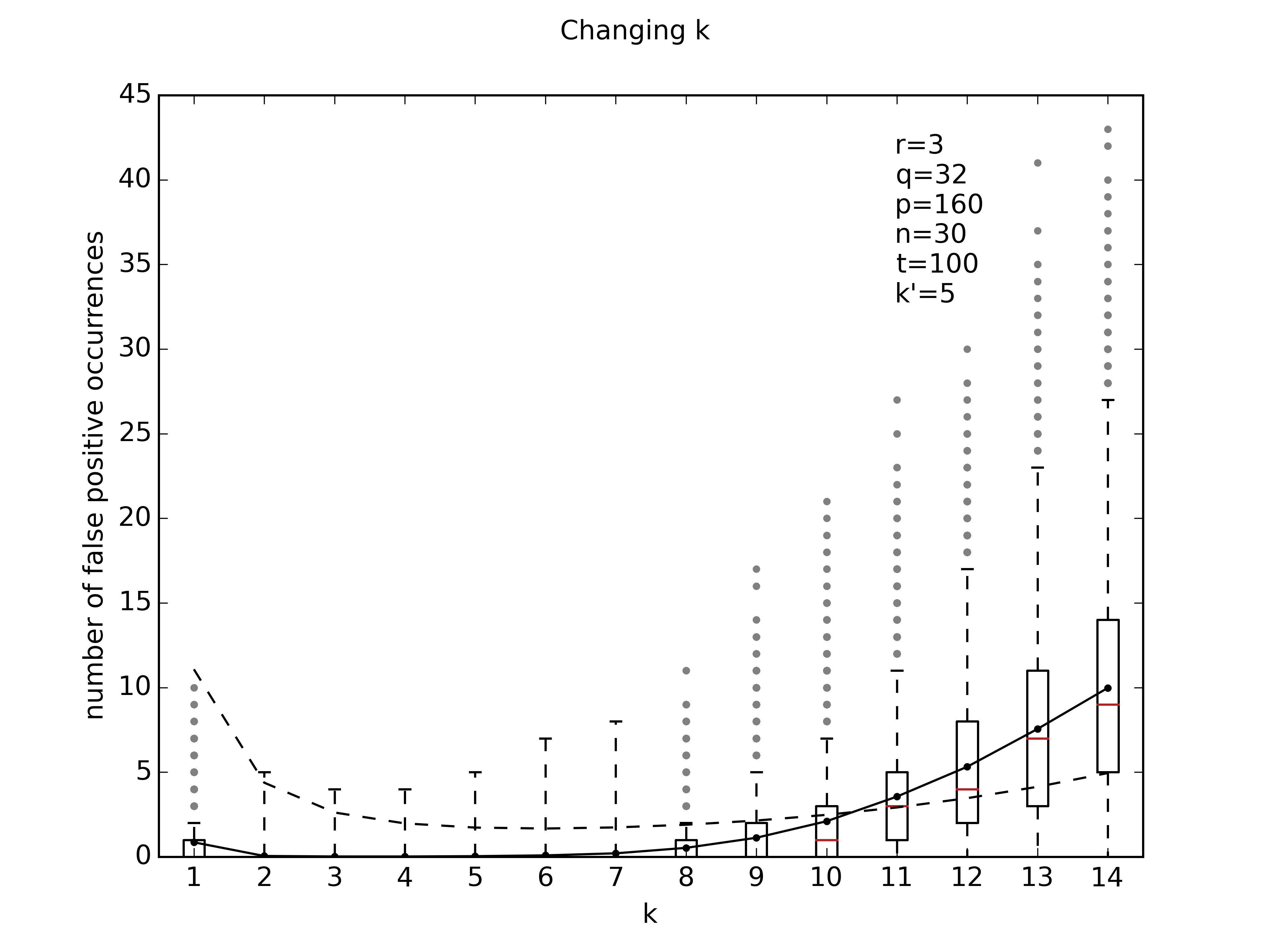}
	\caption{Average number of false positive as a function of the number $k$ of hash functions of the yes-filter. The dotted line represents the average number of false positive of the correspondent BF of size $m=256$.}
	\label{test_fp_ky}
\end{figure}

For the evaluation of false positive performance as a function of the number of hash functions for the yes-filter the average number of false positive occurrences has been computed for $ 1 \leq k \leq 14$. 
The plot in Fig. \ref{test_fp_ky} shows the average number of false positives of the yes-no BF for $10^4$ experiments together with the average number of false positive of the correspondent BF.

\begin{figure}[h]
	\centering
	\includegraphics[width=3.8in]{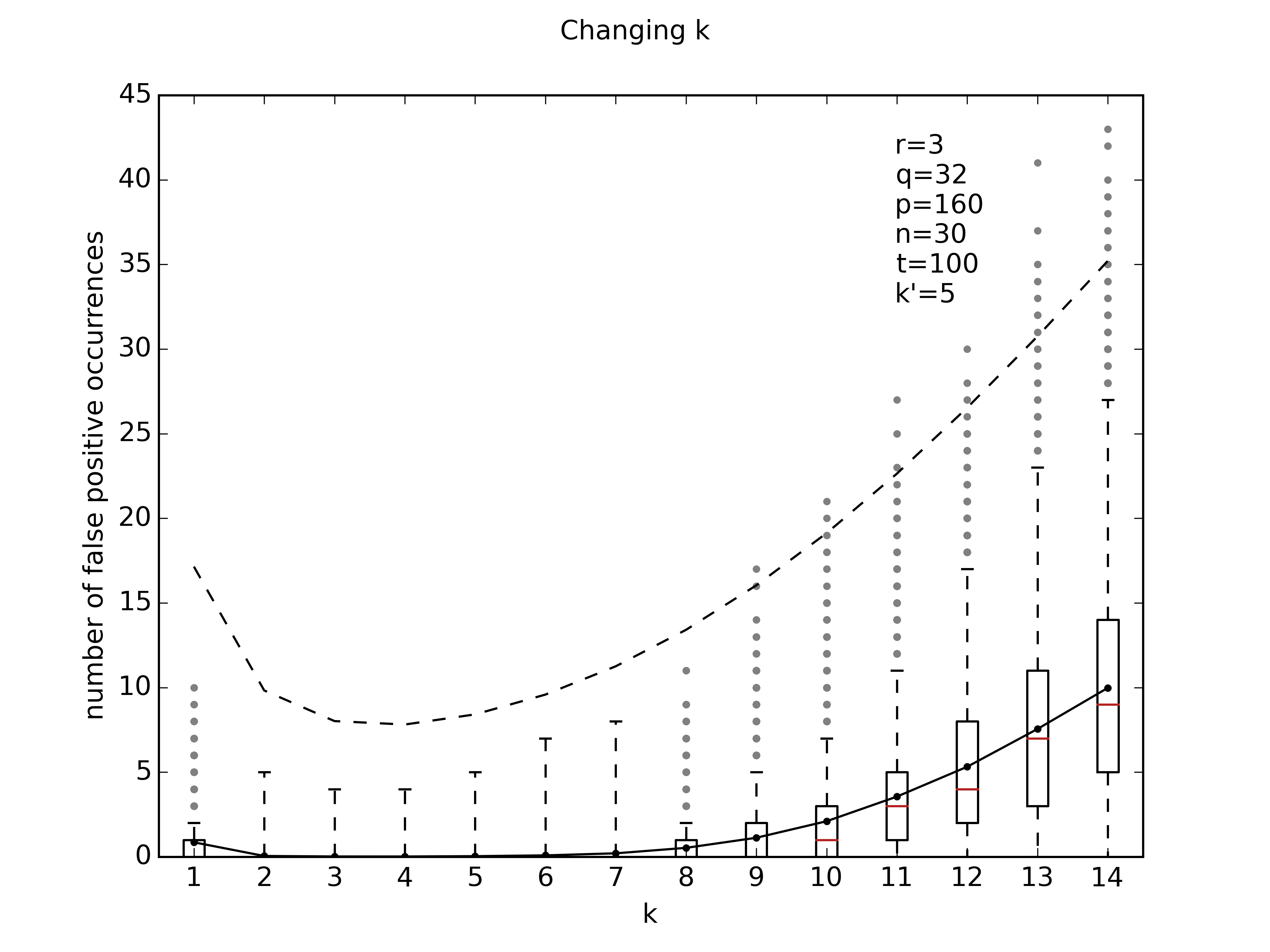}
	\caption{Average number of false positive as a function of the number $k$ of hash functions of the yes-filter. The dotted line represents the average number of false positive of a BF of the same as the yes-filter $m=160$.}
	\label{test_fp_ky_m160}
\end{figure}

The simulations show that the average number of false positive occurrences as a function of the number of hash functions of the yes-filter has a minimum as for the classic BF. Moreover, for small number of hash functions the number of false positive occurrences is much smaller for the yes-no BF while for higher $k$ the BF performs better. This is expected since the yes-filter is smaller in length than the classic BF so as $k$ increases the yes-filter becomes fuller than the BF and the no-filter mechanism which improves the false positive cannot cope any longer. 
Just to analyse the actual reduction in false positive occurrences, we also compare the performance of the yes-no BF with a BF of the same size as the yes-filter. The no-filters mechanism clearly reduces greatly the average number of fasle positive occurrences, as shown in Fig. \ref{test_fp_ky_m160}, following the results of the theoretical analysis of the false positive probability (\ref{PE<PF}).

\begin{figure}[h]
	\centering
	\includegraphics[width=3.8in]{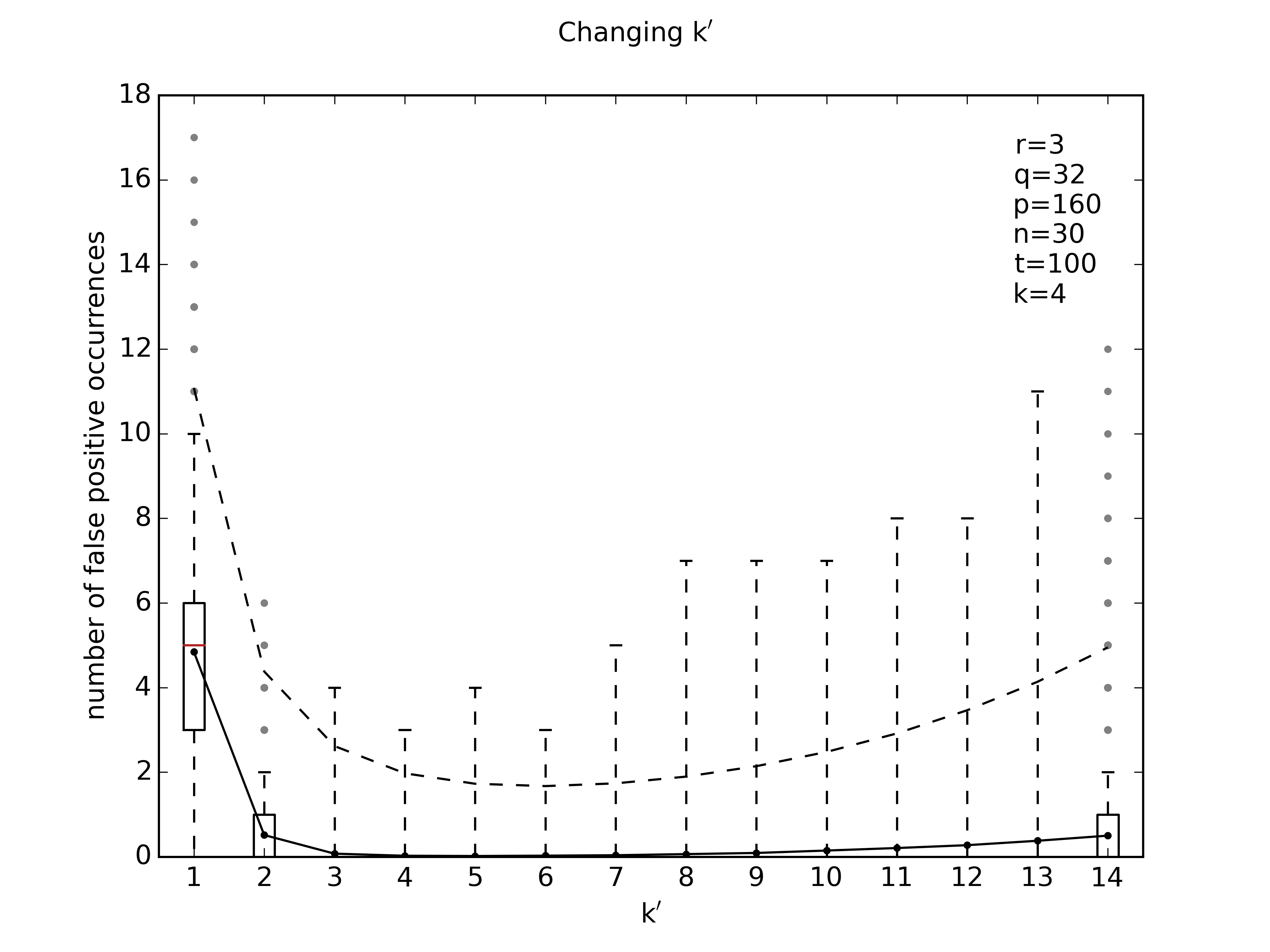}
	\caption{Average number of false positive as a function of the number $k'$ of hash functions of the no-filter. The dotted line represents the average number of false positive of the correspondent BF.}
	\label{test_fp_kn}
\end{figure}

For the evaluation of false positive performance as a function of the number of hash functions for the no-filter the average number of false positive occurrences has been computed for $ 1\leq k' \leq 14$. 
The plot in Fig. \ref{test_fp_kn} shows the average number of false positives of the yes-no BF for $10^4$ experiments together with the average number of false positive of the correspondent BF.

The simulations show that the average number of false positive occurrences as a function of the number of hash functions of the no-filter has a minimum as for the classic BF. Changing the number of hash functions of the no-filter does not influence greatly the number of false positive occurrences and its value remains well below the correspondent value for the classic BF since an appropriate value of $k$ has been chosen for the number of hash function of the yes-filter.  

\subsection{Evaluation of false positive performance as a function of the number of elements to store in the yes-no BF}

\begin{figure}[h]
	\centering
	\includegraphics[width=3.8in]{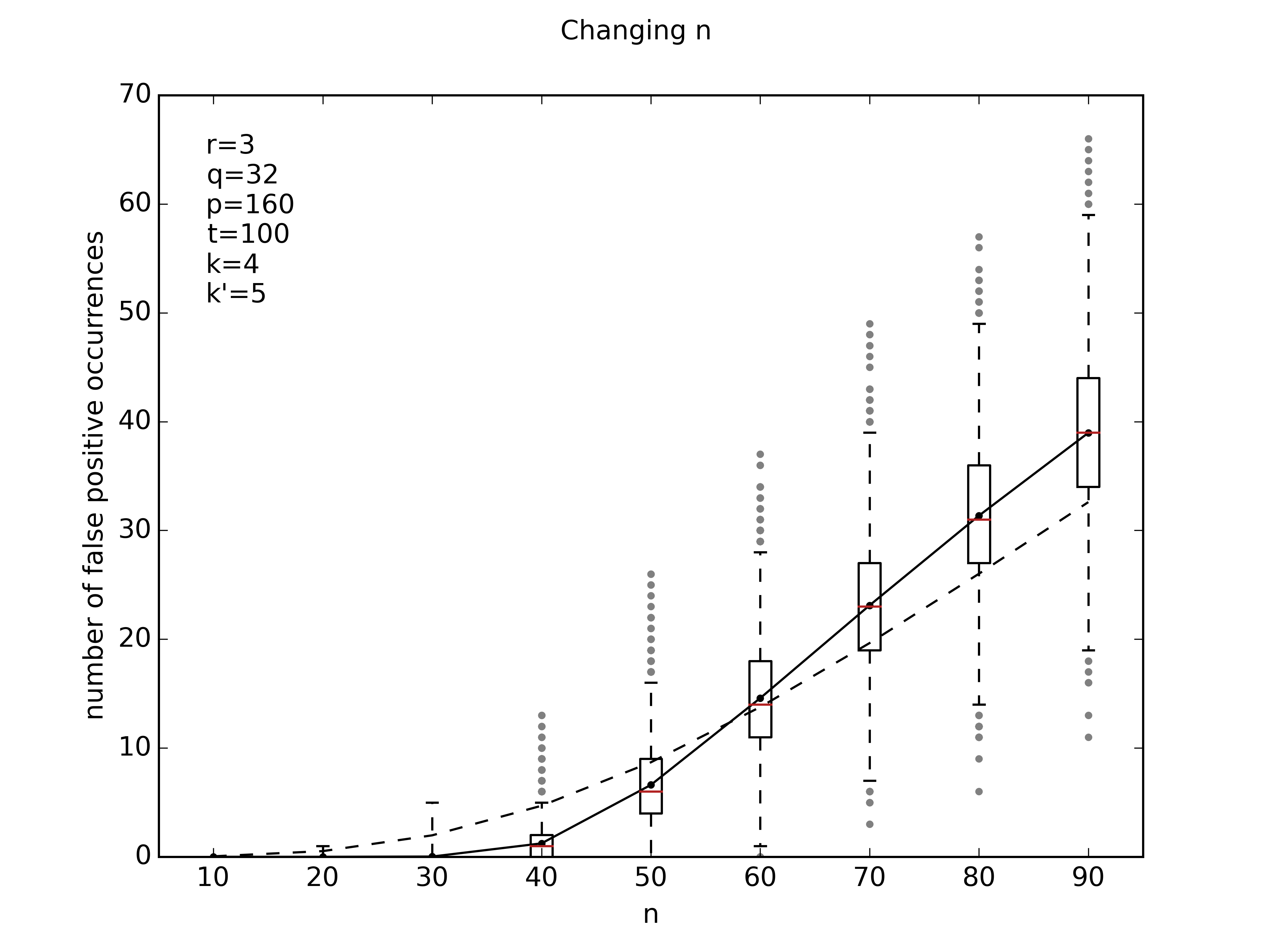}
	\caption{Average number of false positive occurrences as a function of the number $n$ of elements to store in the yes-no BF. The dotted line represents the average number of false positive of the correspondent BF of $m=256$ bits.}
	\label{test_fp_n_m256}
\end{figure}

For the evaluation of false positive performance as a function of the number of the elements to store in the yes-no BF the average number of false positive occurrences has been computed for $ 10\leq n \leq 90$. 
The plot in Fig. \ref{test_fp_n_m256} shows the average number of false positives of the yes-no BF for $10^4$ experiments together with the average number of false positive of the correspondent BF.

\begin{figure}[h]
	\centering
	\includegraphics[width=3.8in]{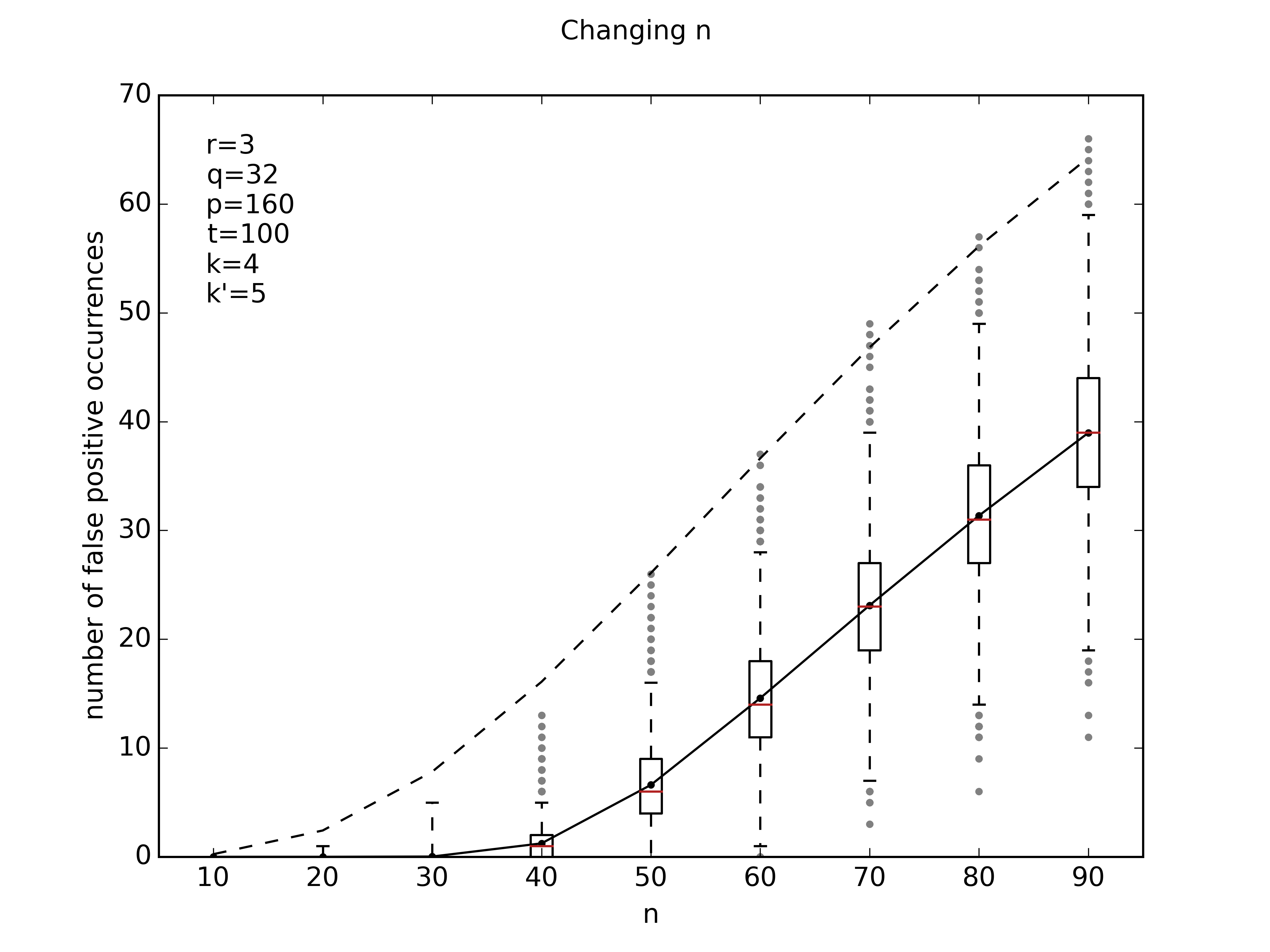}
	\caption{Average number of false positive occurrences as a function of the number $n$ of elements to store in the yes-no BF. The dotted line represents the average number of false positive of the correspondent BF of $m=160$ bits.}
	\label{test_fp_n_m160}
\end{figure}

The simulations show that the average number of false positive occurrences as a function of the number of elements to store in the yes-no BF is an increasing function as for the classic BF. Moreover, for a number of elements lower than around 60 the yes-no BF offers a lower number of false positive occurrences. After this threshold the classic BF performs better on average. This is due to the fact that the number of bits dedicated to store the elements is smaller in the yes-no BF. Just to analyse the actual reduction in false positive occurrences, we compare the performance of the yes-no BF with a BF of the same size as the yes-filter. The yes-no BF offers definitely much less number of false positive occurrences than the classic BF as shown in Fig. \ref{test_fp_n_m160}, following the results of the theoretical analysis of the false positive probability (\ref{PE<PF}).

\subsection{Evaluation of false positive performance as a function of the length of the no-filter}

\begin{figure}[h]
	\centering
	\includegraphics[width=3.8in]{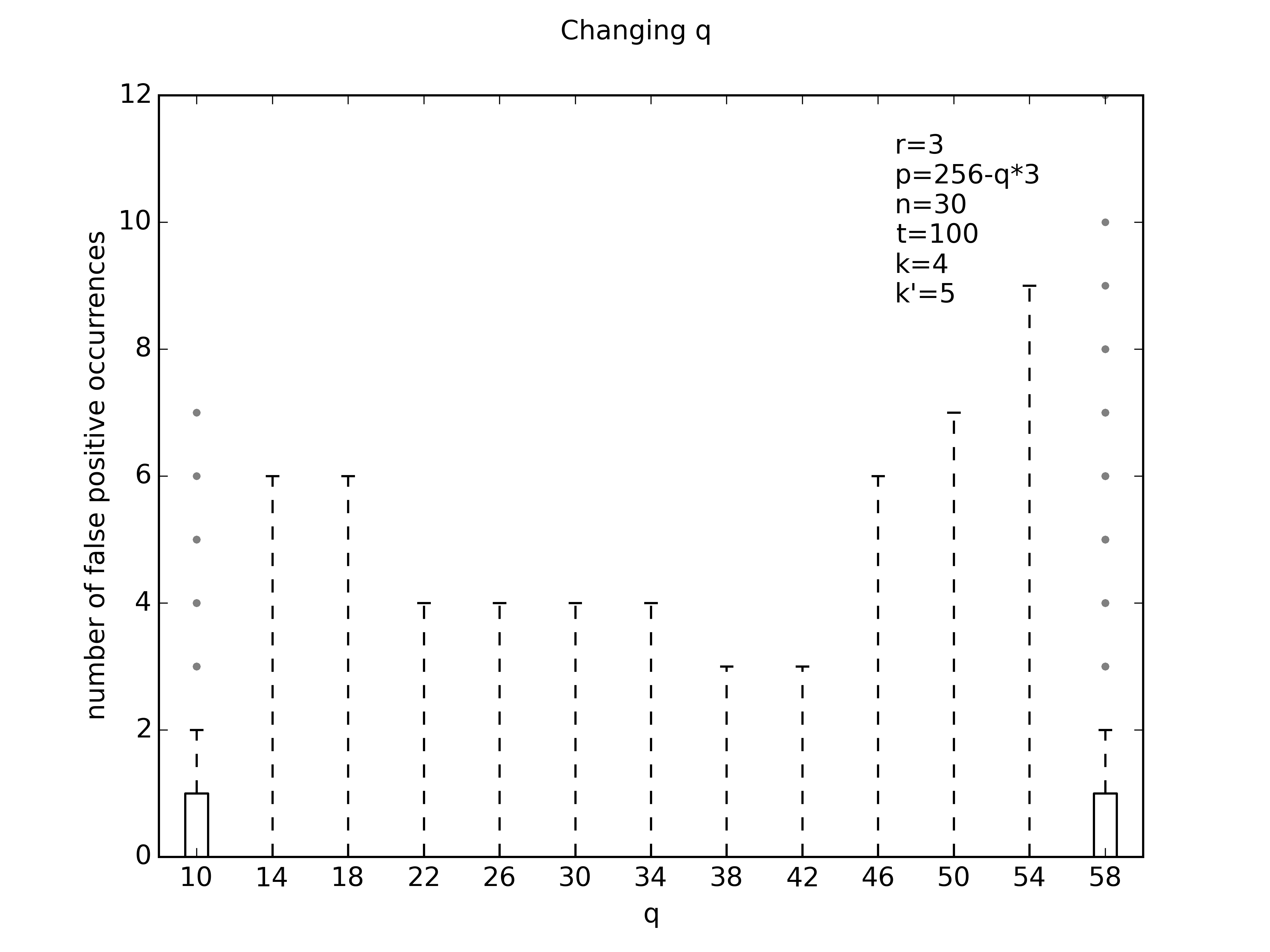}
	\caption{Statistics of number of false positive as a function of the length $q$ of the no-filter.}
	\label{test_fp_q_stat}
\end{figure}

\begin{figure}[h]
	\centering
	\includegraphics[width=3.8in]{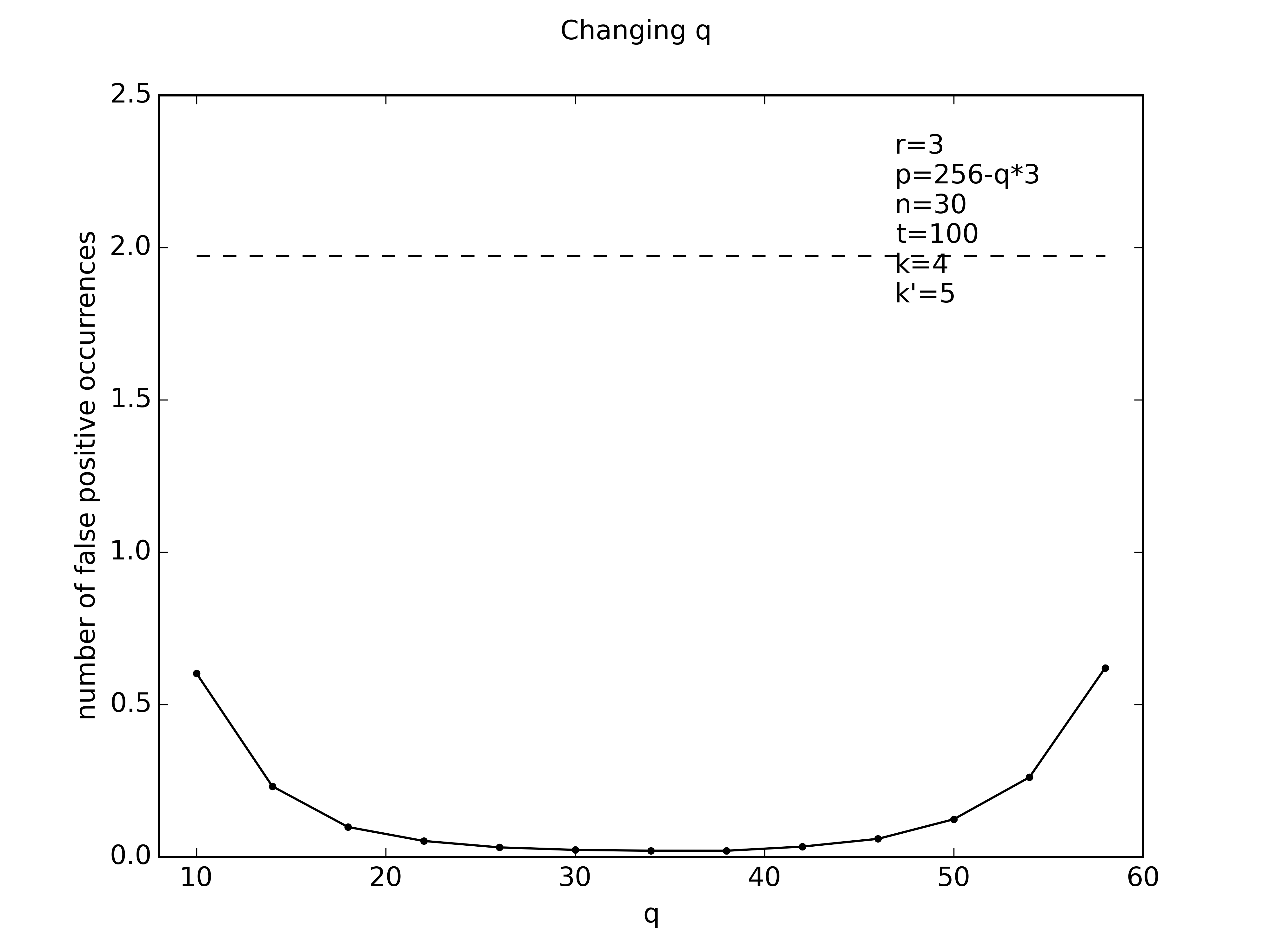}
	\caption{Average number of false positive occurrences as a function of the length $q$ of the no-filter together with the number of false positive occurrences of the classic BF (dotted line).}
	\label{test_fp_q}
\end{figure}

For the evaluation of false positive performance as a function of the length of the no-filter the average number of false positive occurrences has been computed for $ 10\leq \! q \! <  60$. 
The plot in Fig. \ref{test_fp_q_stat} shows the statistic of the average number of false positives of the yes-no BF for $10^4$ experiments together with the average number of false positive of the correspondent BF.
Fig. \ref{test_fp_q} show the averages together with the average number of false positive occurrences for the classic BF.

The simulations show that the average number of false positive occurrences as a function of the length of each no-filters has a minimum value. The number of false positive occurrences is much lower than the correspondent value for the BF since appropriate values has been chosen for the other parameters. The length of each no-filter does not have a strong impact on the overall results.

\subsection{Evaluation of false positive performance as a function of the number of no-filters}

\begin{figure}[h]
	\centering
	\includegraphics[width=3.8in]{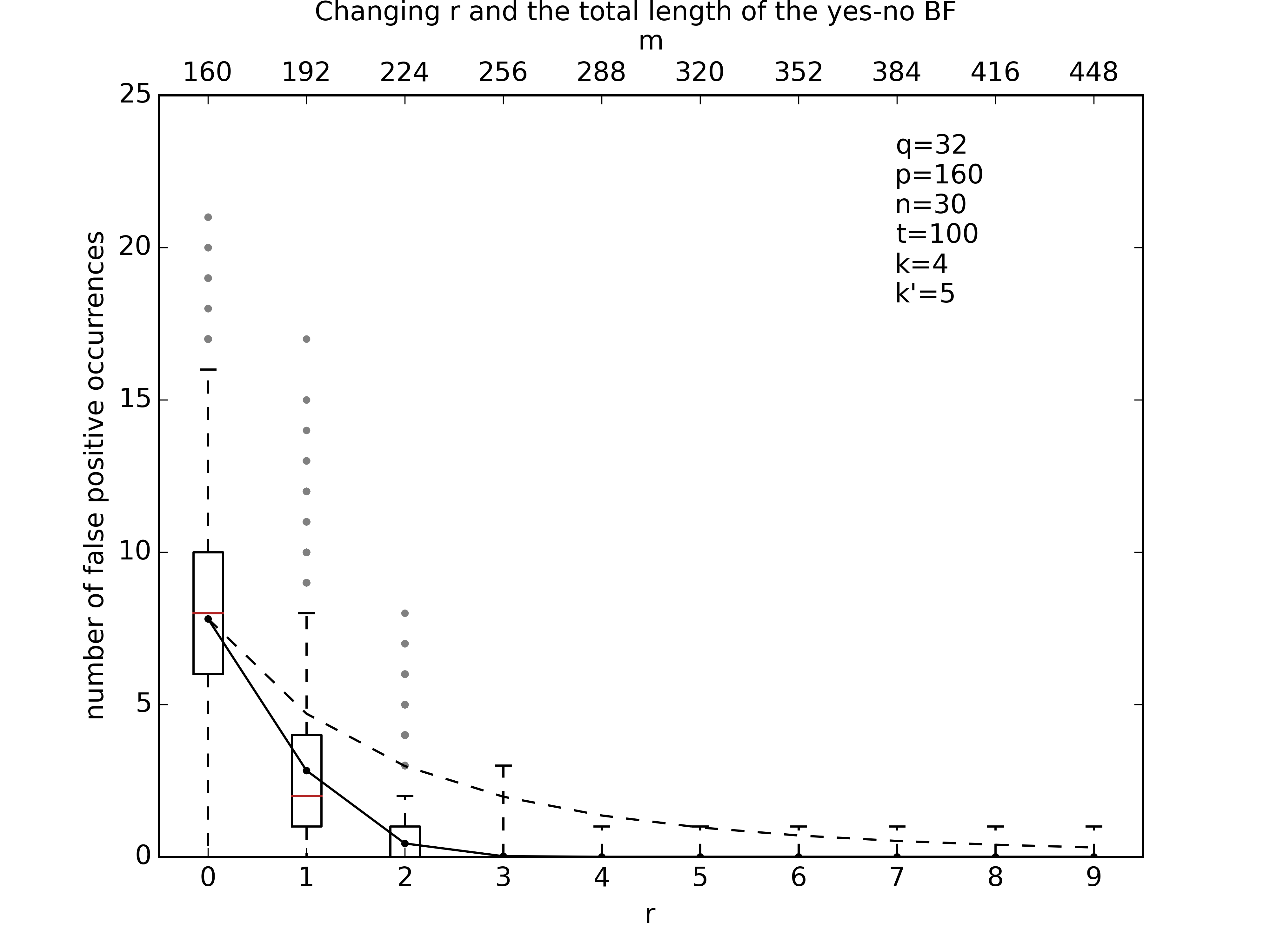}
	\caption{Average number of false positive occurrences as a function of the number $r$ of no-filters, keeping the length of the yes-filter fixed, together with the number of false positive occurrences of the classic BF (dotted line). The length of the yes-no BF is reported on the upper x-axis.}
	\label{test_fp_r_and_total_length}
\end{figure}

For the evaluation of false positive performance as a function of the number of no-filters the average number of false positive occurrences has been computed for $ 1\leq \! r \! \leq 9$. Firstly, we consider the case in which the length $p$ of the yes-filter is kept fixed and consequently the total length of the yes-no BF changes accordingly. Having $p=160$, $q=32$ and $ 1\leq \! r \! \leq 9$, the total length of the yes-no BF is $m=p+qr$ as discussed in Sec. \ref{yesno}.
The plot in Fig. \ref{test_fp_r_and_total_length} shows the average number of false positives of the yes-no BF for $10^4$ experiments together with the average number of false positive of the correspondent BF. The value of $m$ is shown on the upper x-axis.

For $r=0$, the yes-no BF becomes a classic BF where $m=p=160$ while for greater $r$ the yes-no BF offers a lower number of false positives than the classic BF of the same length $m$. This is expected since with more no-filters more false positives can be tracked.  

Secondly, we consider the case in which the length $m$ of the yes-no BF is kept fixed and consequently the length of the yes-filter changes accordingly. Having $m=256$, $q=32$ and $ 1\leq \! r \! \leq 7$, the total length of the yes-filter is $ p=m-qr$ as discussed in Sec. \ref{yesno}.

\begin{figure}[h]
	\centering
	\includegraphics[width=3.8in]{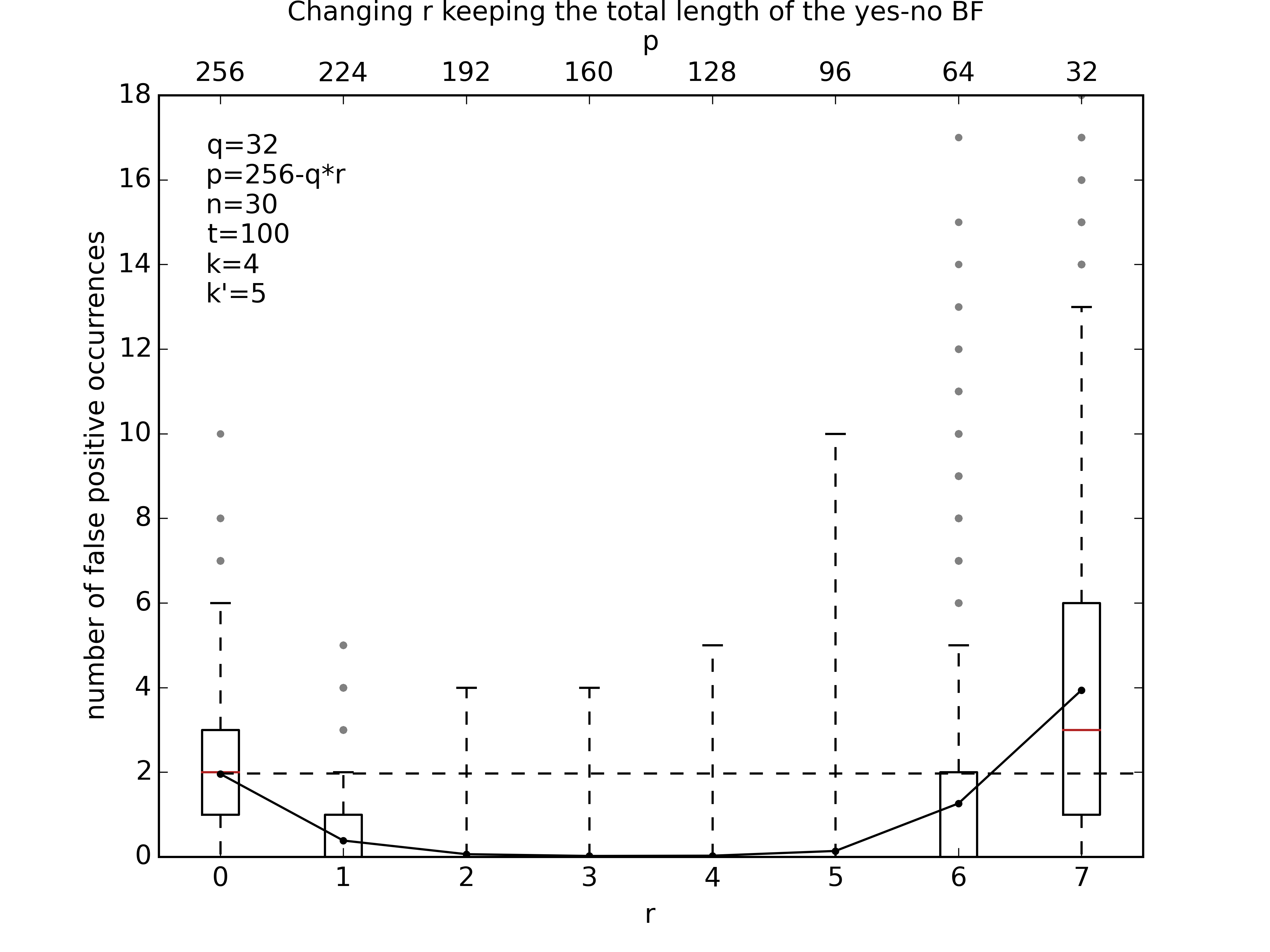}
	\caption{Average number of false positive occurrences as a function of the number $r$ of no-filters keeping the total length of the yes-no BF fixed, together with the number of false positive occurrences of the classic BF (dotted line).The length of the yes-filter is reported on the upper x-axis.}
	\label{test_fp_r}
\end{figure}

Fig. \ref{test_fp_r} shows the averages together with the average number of false positive occurrences for the classic BF. The value of $p$ is shown on the top x-axis.

Again, when $r=0$ the yes-no Bf becomes the classic BF, but as $r$ increases the number of false positive occurrences decreases. However, the function has a minimum since because the total length of the yes-no BF is kept constant, the length $p$ of the yes-filter will become smaller and smaller generating too many false positives that the yes-no BF will be not be able to cope with.

%%%%%%%%%%%%%%%%%%%%%%%%%%%%

According to this study, we have chosen appropriate values of these parameters for the application described in Sec. \ref{yntopo}.
%In the description of our experiments below, we suggest appropriate values of these parameters which have been determined experimentally.

\section{Algorithm complexity}
The processing required to form the yes-no BF can be quantified using the big O notation. 
Regarding the construction of the yes-filter, the procedure is exactly the same as for the classic BF so we do not consider this in this analysis since we want to quantify how much more processing is required with respect to the classic BF. 
The complexity comes from the no-filters formation and we analyse the case in which false negatives need to be avoided.

We assume that the operations to form and to query a BF can be parallelised. 

First of all the set $F$, the set of the elements generating false positives has to be built. The complexity of this operation can be quantified as $O(|T|)$, where $|T|$ is the cardinality of the set $T$. This is because each element in the set $T$ has to be checked.  

Given $F$, if the false negatives can be tolerated then a greedy algorithm can be used to choose in which no-filter of the set $S$ the no-filter of the false positive can be stored. The time complexity of a greedy algorithm can be generally estimated as linear in the number of false positives $O(|F|)$.

If the false negatives have to be avoided then an additional processing is required. In the worst case scenario, the time complexity of the algorithm can be quantified as $O(r \,|F||S|)$ since the no-filter of each element in $S$ has to be compared with the no-filter in consideration for each false positive and in the worst case scenario this has to be repeated for all the no-filters. 

Since the operation involved can be implemented as logical \verb|and|, \verb|or| and comparison we can conclude that the processing overhead is only slightly more than for the classic BF at the formation stage. 

At the querying stage, the processing required by the yes-no BF is exactly the same as for the BF, since the membership query can be implemented as a logical \verb|and| and comparison for both the data structure. BF and yes-no BF can answer membership queries in O(1) time.

\section{The yes-no BF for packet forwarding}
\begin{figure}[h]
	\centering
	\includegraphics[width=3.4in]{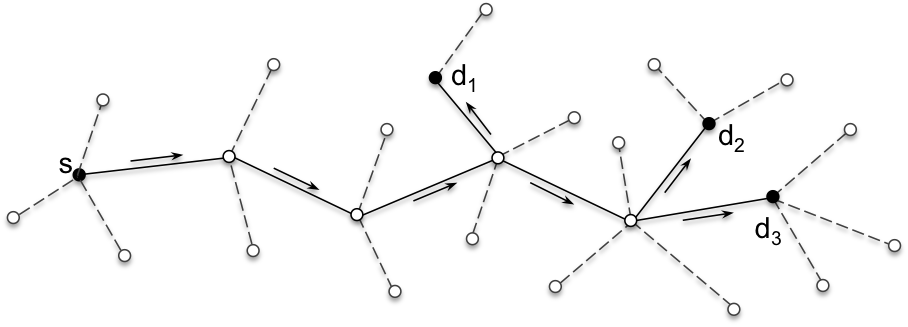}
	\caption{Schematic representation of a path with multiple destinations.}
	\label{MulticastTree1}
\end{figure}

The formulation of the yes-no BF has been inspired by the recent application of BFs in the novel multicast forwarding fabric \cite{jokela} of the PSIRP/PURSUIT information centric architecture.  However, structures like the BF and the yes-no BF can be used generally for packet forwarding when the path is known. These structures are used as an encoding to identify paths (with one or more destinations) and links between nodes. The yes-no BF representing an element described in Sec. \ref{yesnoele} is used to encode a link, where $e$ would be a link descriptor. The yes-no BF representing a set described in Sec. \ref{yesnoset} is used to encode the path. Thus, the set of all links is $U$, the set of links along the path is $S$, and the set of all the outgoing links from the nodes along the path is $T$. Fig. \ref{MulticastTree1} shows a schematic representation of a path of multiple destinations where link belonging to the set $S$ and the links belonging to the set $T$ are differently depicted. 

The yes-no BF, encoding the path, is placed in the packet header. Note that slightly more processing is required to construct the yes-no BF in comparison with the construction of the BF: the false positive occurrences for the specific path have to be evaluated and the no-filters have to be constructed.
At a node, the yes-no BF of each outgoing link is stored. When the packet arrives at the node, the membership query described in Sec. \ref{yesnoquery} is performed for each of the outgoing links. Note that the membership query can be implemented with a logical \verb|and| operation and a comparison, exactly like the membership query for the BF.  Therefore, we can conclude that the yes-no BF requires a bit more processing at the stage of its formation but the same processing for membership queries as the BF, so that the forwarding operation can be implemented in line speed \cite{jokela}.  A query may result in a false positive, which translates, in our forwarding scheme, into extra traffic in the network. The match is possible along more than one outgoing link per node so that multicast turns out to be the natural communication paradigm.  The formation of the yes-no BF is performed by an entity that has knowledge of the network topology, this could a centralised entity \cite{fotiou}.

\section{Evaluation of the yes-no BF performance in realistic topologies}\label{yntopo}
\begin{figure}
  \centering
  \includegraphics[width=3.0in]{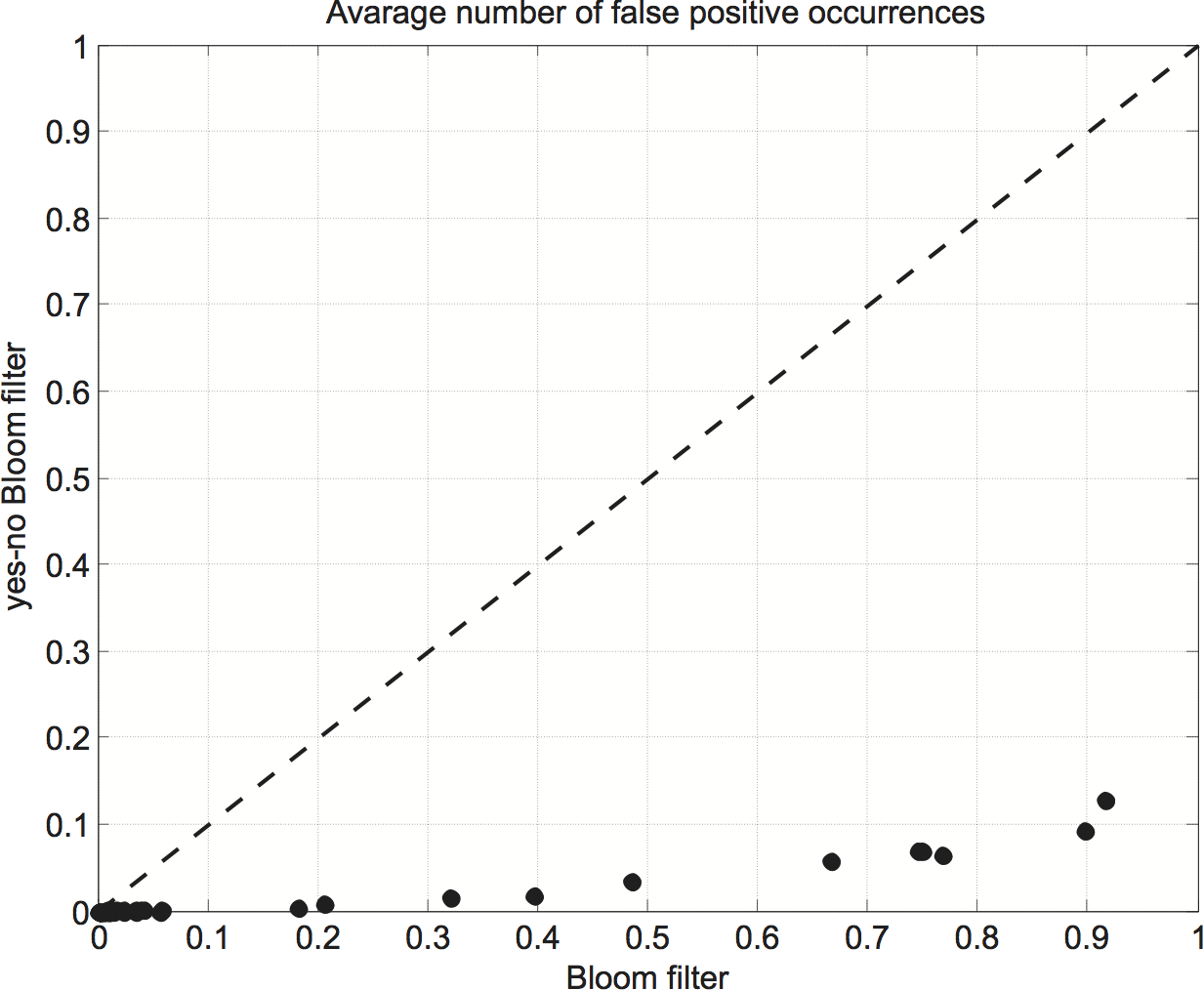}
  \caption{The expected number of false positive occurrences of the yes-no BF against the BF.}
  \label{Fig1}
\end{figure}

Although the evaluation of the yes-no BF performance could be carried out in artificially generated topologies, we consider realistic topologies and we compare the false positive rates of the yes-no BF and of the BF of the same size. The network topologies used in the simulation are a collection from the Network Topology Zoo containing 261 network topologies \cite{knight}. In each of these 261 networks, we select a long (relative to the topology) realistic forwarding path.  The forwarding path is represented as the set $S$ containing all the links along the path, whereas all the other links adjacent to the path, which are obviously not part of the path, are collected in the set $T$. Second order false positives are not considered since the probability is multiplicative and it can be considered negligible. For each of the elements of the sets $S$ and $T$ we consider 1000 random allocations of the position of the 1s for the yes-filter and the no-filter to simulate the hashing. We build the yes-no BF for all the elements of $S$ and $T$ and calculate the number of false positives as the average over the number of false positive occurrences for the 1000 allocations.
The parameters defining the yes-no BF used in the simulation are listed below:
\begin{itemize}
	\item The total length of the yes-no BF and of the BF is $m=256$ bits.
	\item The length of the yes-filter is $p=192$ bits.
	\item The number of no-filters are $r=2$.
	\item The length of each no-filter is $r=32$ bits.
	\item The number $k$ of hash functions used for the yes-filter is $k=4$.
	\item The number $k'$ of hash functions used for the no-filters is $k'=3$.
	\item The number $k_{\texttt{BF}}$ of hash functions for the classic BF is $k_{\texttt{BF}}=6$.
\end{itemize}
These parameters have been chosen according to the analysis performed in Sec. \ref{evalFP}. The total length, $m$ matches typical lengths used in other studies and is comparable with the length of the IPv6 header address fields. We do not claim that this is the optimal parameter set but the aim of this paper is to show that the yes-no BF offers better false positive performance than the classic BF.
%\subsection{Results}
Fig. \ref{Fig1} shows the expected number of false positive occurrences of the yes-no BF against the expected number of false positive occurrences of the BF, where it can be easily seen that the yes-no BF offers a lower number of false positive occurrences.
For example, the top right point corresponds to a 28-hop path in the network from the TataNld topology \cite{knight}. If the BF is used for forwarding along this path, one can expect, approximately, 0.92 false positive occurrences; but if the yes-no BF is used, one can expect, approximately, 0.13 false positives. The meaning of this is that if we use BFs, packets will almost certainly be diverted to a wrong direction approximately at one point during their travel; however, if we use yes-no BFs, this will only happen approximately to one out of eight possible yes-no BFs of the links in the network.

% (or yes-no BF of the elements in $U$).
%\begin{figure}
%  \centering
%  \includegraphics[width=2.5in]{FigOHk1}
%  \caption{The false positive rate of the optihash with one hash per link and for $N=32768$ in comparison with the false positive rate of the Bloom filter with $m=241$ $k=1$ and with $m=256$ $k=7$.}
%  \label{FigOHk1}
%\end{figure}
Note there are many very small networks for which neither the BF nor, of course, the yes-no BF produce any false positives. They are not interesting for our comparison. On the diagram, they all are represented by the point with the coordinates 0,0.
Furthermore, there are two comparatively large topologies, in which both the BFs and the yes-no BFs produce a relatively large number of false positives, and, therefore, perhaps one would want to consider using a different forwarding model. 
%We note that the yes-no BF still performs noticeably better than the BF, as reported in table \ref{table_example}.
%\begin{table}[tb]
%\renewcommand{\arraystretch}{1.6}
%\caption{Average number of false positive occurrences for the worst case topologies over 1000 allocations}
%\label{table_example}
%\centering
%\begin{tabular}{|l|c|c|c|}
%\hline
%Topology & Path length & BF FP & Yes-No BF FP \\
%\hline\hline
% UsCarrier & 35 & 1.95 & 0.48 \\
% \hline
% Kdl & 58 & 15.87 & 11.27 \\
%\hline
%\end{tabular}
%\end{table}

\begin{figure}[tb]
  \centering
  \includegraphics[width=3.2in]{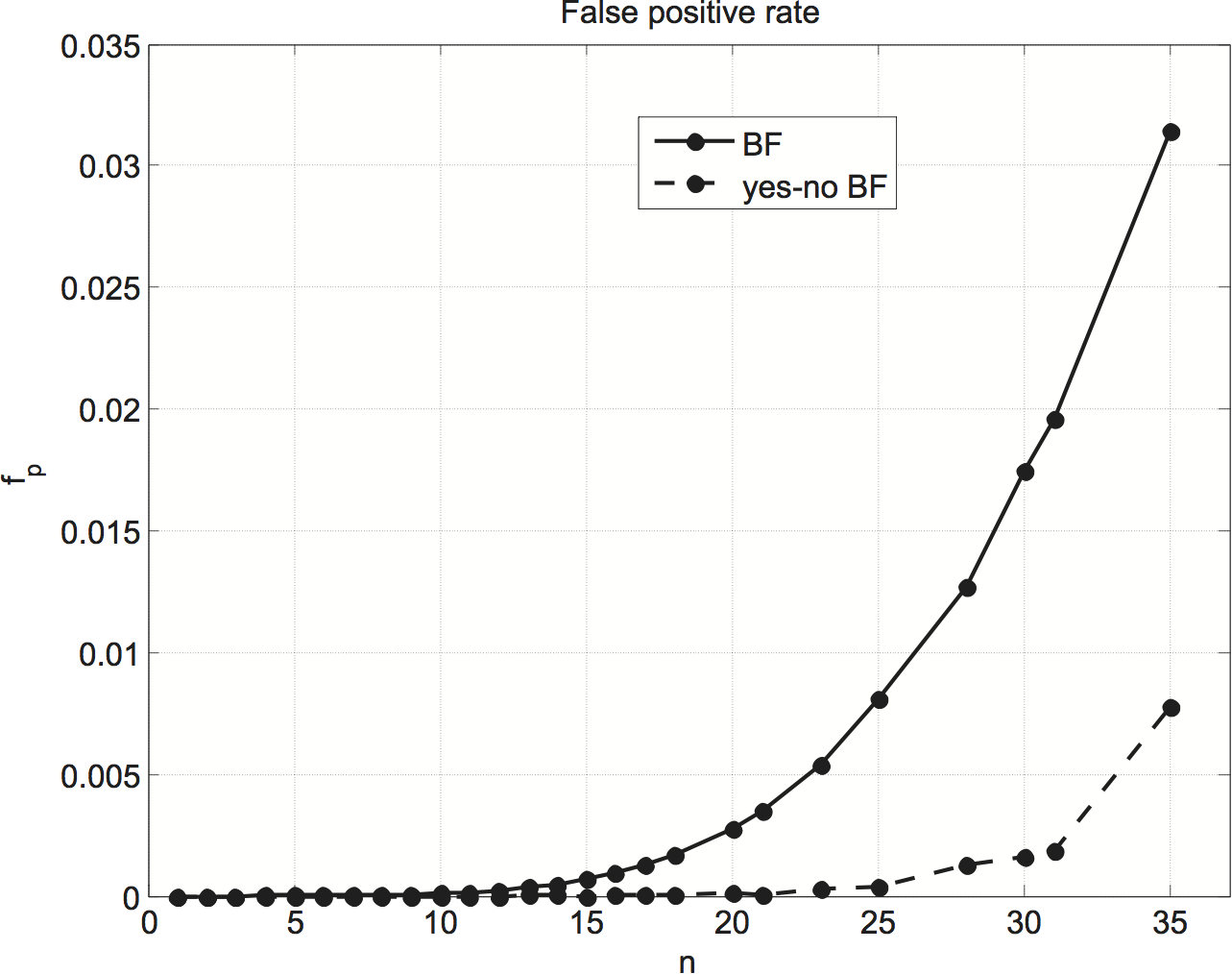}
  \caption{The false positive rate of the yes-no BF and of the BF against the number $n$ of links along the path.}
  \label{Fig2}
\end{figure}

Fig. \ref{Fig2} shows the false positive rate against the number of links along the path of the BF and the yes-no BF. The result has been obtained averaging the false positive rate of paths on different topologies having the same length.
The false positive rate of the yes-no BF is consistently lower than the false positive rate of the BF.

\begin{figure}
	\centering
	\includegraphics[width=3.0in]{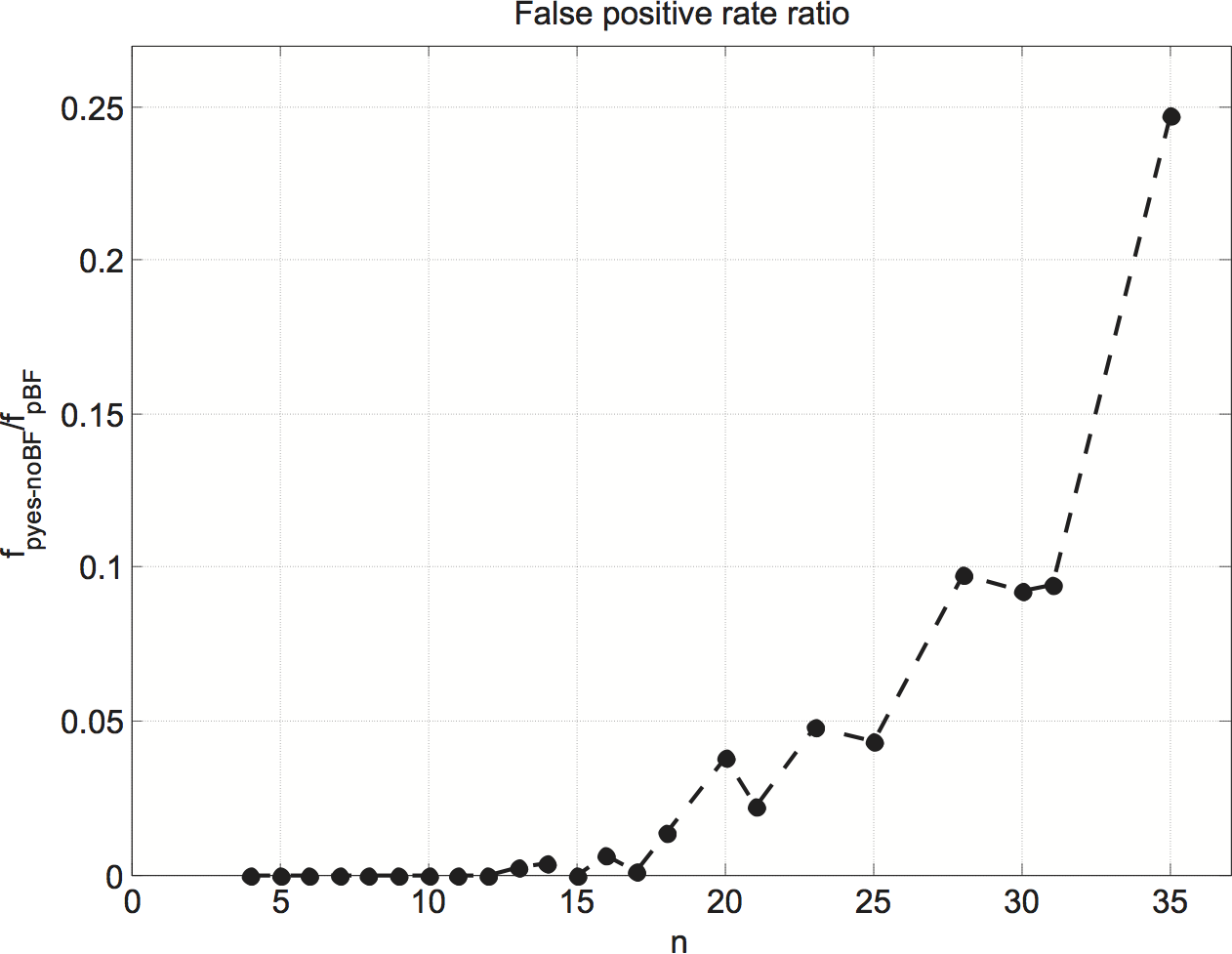}
	\caption{The ratio between the false positive rate of the yes-no BF and of the BF as a function of the number of links encoded in the data structure.}
	\label{Fig3}
\end{figure}

Fig. \ref{Fig3} shows the ratio of the false positive rate of the yes-no BF and of the BF $f_{yn}/f_{BF}$ as a function of the number of links (elements) encoded in the data structure, up to $n=35$ links where the yes-no BF offers a false positive rate which only a quarter of the correspondent false positive rate of the BF.  
%\begin{figure}
%  \centering
%  \includegraphics[width=2.5in]{Fig3}
%  \caption{The ratio of the false positive rate of the yes-no BF and of the BF against the number $n$ of links along the path. }
%  \label{Fig3}
%\end{figure}
\section{Conclusions}
We have proposed a novel way of representing sets based on BFs, which we call the yes-no BF. It requires slightly more processing at the stage of its formation and the same processing for membership queries as the classic BF, but offers a considerably smaller number of false positives and no false negatives. We have shown using computational experiments that the yes-no BF outperforms the classic BF in the scenario of packet forwarding with in-packet path encoding as introduced in the information-centric architecture PSIRP/PURSUIT. However, the structure is general enough to be used in a wide variety of applications. Moreover, the yes-no BF construction is dynamic and allows a choice of heuristics and optimization algorithms, as demonstrated in \cite{yang}.

\section*{Acknowledgment}
%This work has been carried out with the support of
%the EPSRC and British Telecom(BT) through an EPSRC CASE
%award studentship and with the support of an EU FP7 project PURSUIT
%under Grant FP7-INFSO-ICT 257217.

\bibliographystyle{IEEEtran}
\bibliography{biblioPSIRP}

\begin{thebibliography}{10}
\providecommand{\url}[1]{#1}
\csname url@samestyle\endcsname
\providecommand{\newblock}{\relax}
\providecommand{\bibinfo}[2]{#2}
\providecommand{\BIBentrySTDinterwordspacing}{\spaceskip=0pt\relax}
\providecommand{\BIBentryALTinterwordstretchfactor}{4}
\providecommand{\BIBentryALTinterwordspacing}{\spaceskip=\fontdimen2\font plus
\BIBentryALTinterwordstretchfactor\fontdimen3\font minus
  \fontdimen4\font\relax}
\providecommand{\BIBforeignlanguage}[2]{{%
\expandafter\ifx\csname l@#1\endcsname\relax
\typeout{** WARNING: IEEEtran.bst: No hyphenation pattern has been}%
\typeout{** loaded for the language `#1'. Using the pattern for}%
\typeout{** the default language instead.}%
\else
\language=\csname l@#1\endcsname
\fi
#2}}
\providecommand{\BIBdecl}{\relax}
\BIBdecl

\bibitem{bloom}
B.~Bloom, ``Space/time trade-offs in hash coding with allowable errors,''
  \emph{Commun. ACM}, vol. 13, no.7, pp. 422--426, 1970.

\bibitem{broder}
A.~Broder and M.~Mitzenmacher, ``Network applications of {B}loom filters: A
  survey,'' \emph{Internet Mathematics}, vol. 1, no.4, pp. 485--509, 2004.

\bibitem{tarkoma}
S.~Takoma, C.~Rothenberg, and E.~Lagerspetz, ``Theory and practice of {B}loom
  filters for distributed systems,'' \emph{IEEE Communications Surveys and
  Tutorials}, vol. 14(1), pp. 131--155, 2012.

\bibitem{donnet}
B.~Donnet, B.~Baynat, and T.~Friedman, ``Retouched {B}loom filters: Allowing
  networked applications to trade off selected false positives against false
  negatives.''\hskip 1em plus 0.5em minus 0.4em\relax Lisboa, Portugal: CoNEXT
  2006, Dec 2006.

\bibitem{rothenberg2010:deletable}
C.~Rothenberg, C.~Macapuna, F.~Verdi, and F.~Magalhaes, ``The deletable {B}loom
  filter: a new member of the {B}loom family,'' \emph{IEEE Communications
  Letters}, vol. 14,no.6, pp. 557--559, 2010.

\bibitem{fan}
L.~Fan, P.~Cao, J.~Almeida, and A.~Broder, ``Summary cache: A scalable
  wide-area web cache sharing protocol,'' \emph{IEEE/ACM Transaction on
  Networking}, vol. 8, no.3, pp. 281--293, 2000.

\bibitem{ficara}
D.~Ficara, S.~Giordano, G.~Procissi, and F.~Vitucci, ``Multilayer compressed
  counting {B}loom filters,'' in \emph{Proceedings of the 27$^{th}$ Annual
  Joint Conference of the IEEE Computer and Communications Societies
  (INFOCOM)}, 2008, pp. 311--315.

\bibitem{almeida:scalable}
P.~Almeida, C.~Baquero, N.~Preguica, and D.~Hutchison, ``Scalable {B}loom
  filter,'' \emph{Information Processing Letters}, vol. 101, no.6, pp.
  255--261, 2007.

\bibitem{lim15}
H.~Lim, J.~Lee, and C.~Yim, ``Complement {B}loom filter for identifying true
  positiveness of a {B}loom filter,'' \emph{IEEE Communication Letters}, vol.
  19, no.11, pp. 1905--1908, 2015.

\bibitem{carrea}
L.~Carrea, A.~Vernitski, and M.~Reed, ``Optimized hash for network path
  encoding with minimized false positives,'' \emph{Computer Networks}, vol.~58,
  pp. 180--191, 2014.

\bibitem{jokela}
P.~Jokela, A.~Zahemszky, C.~E. Rothenberg, S.~Arianfar, and P.~Nikander,
  ``{L}{I}{P}{S}{I}{N}: Line speed publish/subscribe inter-networking.''\hskip
  1em plus 0.5em minus 0.4em\relax Barcelona, Spain: ACM SIGCOMM '09, Aug 2009.

\bibitem{fotiou}
N.~Fotiou, G.~Polyzos, and D.~Trossen, ``Illustrating a publish/subscribe
  internet architecture,'' \emph{Journal on Telecommunication Systems -
  Springer}, vol. 51, no. 4, pp. 233--245, 2012.

\bibitem{trossen:pekka}
D.~Trossen, J.~Riihijärvi, P.~Nikander, P.~Jokela, J.~Kjällmand, and
  J.~Rajahalme, ``Designing, implementing and evaluating a new internetworking
  architecture,'' \emph{Computer Communications}, vol. 35, no.17, pp.
  2069--2081, 2012.

\bibitem{zahemszkyMPSS}
A.~Zahemszky, P.~Jokela, M.~Sarela, S.~Ruponen, J.~Kempf, and P.~Nikander,
  ``{M}{P}{S}{S}: Multiprotocol stateless switching.''\hskip 1em plus 0.5em
  minus 0.4em\relax San Diego CA USA: 13th IEEE Global Internet Symposium 2010,
  Mar 2010.

\bibitem{bose}
P.~Bose, H.~Guo, E.~Kranakis, A.~Maheshwari, P.~Morin, J.~Morrison, M.~Smid,
  and Y.~Tang, ``On the false positive rate of {B}loom filters,'' \emph{Inf.
  Process. Lett.}, vol. 108, no.4, pp. 210--213, 2008.

\bibitem{kirsch}
A.~Kirsch and M.~Mitzenmacher, ``Less hashing, same performance: Building a
  better {B}loom filter,'' \emph{Random Structures and Algorithms}, vol. 32,
  no.2, pp. 187--218, 2008.

\bibitem{lim}
H.~Lim, N.~Lee, J.~Lee, and C.~Yim, ``Reducing false positives of a {B}loom
  filter using cross-checking {B}loom filters,'' \emph{Appl. Math. Inf. Sci.},
  vol. 8, no.4, pp. 1865--1877, 2014.

\bibitem{knight}
S.~Knight, H.~Nguyen, N.~Falkner, R.~Bowden, and M.~Roughan, ``The {I}nternet
  topology zoo,'' \emph{IEEE Journal on Selected Areas in Communications}, vol.
  29, no.9, pp. 1765--1775, 2011.

\bibitem{yang}
X.~Yang, A.~Vernitski, and L.~Carrea, ``An approximate dynamic programming
  approach for improving accuracy of lossy data compression by {B}loom
  filters,'' \emph{European Journal of Operational Research - IN PRESS}.

\end{thebibliography}






% \end{thebibliography}

\end{document}